\documentclass[aps,prc,twocolumn,showpacs,superscriptaddress,groupedaddress]%
{revtex4-1}

\usepackage[T1]{fontenc}
\usepackage[utf8]{inputenc}
\usepackage[english]{babel}

\usepackage{graphicx}
\usepackage[%
 multi-part-units = single,
 separate-uncertainty = true,
]{siunitx}
\usepackage{tabularx}
\usepackage{booktabs}
\usepackage{amssymb}
\usepackage{xspace}

\usepackage{caption,subcaption}
\usepackage{ragged2e}
\DeclareCaptionJustification{justified}{\justifying}
\newcolumntype{R}{>{\raggedleft\arraybackslash}X}%
\newcolumntype{C}{>{\centering\arraybackslash}X}%
\newcolumntype{L}{>{\raggedright\arraybackslash}X}%

\usepackage[version=3]{mhchem}

\newcommand{\eras}[2]{\ {^{#1}_{#2}}}

\newcommand{\hades}{HADES\xspace}

\newcommand{\tcms}{centre-of-mass\xspace}
\newcommand{\xsec}{cross section\xspace}
\newcommand{\xsecs}{cross sections\xspace}

\newcommand{\phasespace}{phase space\xspace}

\newcommand{\eq}{\begin{equation}}
\newcommand{\eeq}{\end{equation}}
\newcommand{\ud}{\mathrm{d}}
\newcommand{\dEdx}{$\ud E$/$\ud x$\xspace}

\newcommand{\hyp}{\ce{Y}}

\newcommand{\Kstar}{\ce{K}^*}
\newcommand{\prot}{\ce{p}}
\newcommand{\neut}{\ce{n}}

\newcommand{\piz}{\pi^{0}}
\newcommand{\pim}{\pi^{-}}
\newcommand{\pip}{\pi^{+}}

\newcommand{\Kzs}{\ce{K^0_S}}
\newcommand{\Kz}{\ce{K^0}}

\newcommand{\Kp}{\ce{K^+}}
\newcommand{\Dp}{{\Delta^{+}}}
\newcommand{\Dpp}{{\Delta^{++}}}
\newcommand{\Szero}{{\Sigma^{0}}}
\newcommand{\Splus}{{\Sigma^{+}}}
\newcommand{\Sstar}{{\Sigma(1385)}}

\newcommand{\Sstarz}{{\Sigma(1385)^0}}
\newcommand{\Sstarp}{{\Sigma(1385)^+}}
\newcommand{\Lzero}{{\Lambda}}
\newcommand{\Lstar}{{\Lambda(1405)}}
\newcommand{\Lstard}{{\Lambda(1520)}}
\newcommand{\Nstar}{{\text{N}^{*}}}
\newcommand{\pKL}{{\prot\Kp\Lzero}}
\newcommand{\pKS}{{\prot\Kp\Szero}}
\newcommand{\pp}{\ce{pp}\xspace}
\newcommand{\cms}{\mathrm{cms}}

\newcommand{\gev}{\giga\electronvolt}
\newcommand{\gevc}{\giga\electronvolt\per c}

\newcommand{\mev}{\mega\electronvolt}
\newcommand{\mevc}{\mega\electronvolt\per c}
\newcommand{\mevsc}{\mega\electronvolt\per c\squared}

\newcommand{\ptcm}{$p_\mathrm{t}$\xspace}
\newcommand{\ycm}{$y_\cms$\xspace}
\newcommand{\pty}{$p_\mathrm{t}$-$y_\cms$\xspace}

\newcommand{\pcm}{$p_\cms$\xspace}
\newcommand{\cthcm}{$\cos\theta_\cms$\xspace}
\newcommand{\pcth}{$p_\cms$-$\cos\theta_\cms$\xspace}

\newcommand{\mptcm}{p_\mathrm{t}}
\newcommand{\mycm}{y_\cms}
\newcommand{\mpcm}{p_\cms}
\newcommand{\mcthcm}{\cos\theta_\cms}

\newcommand{\racs}{\sqrt{s}}
\newcommand{\hracs}{\racs = \SI{2.7}{\gev}}

\DeclareSIUnit\pion{\pi}

\usepackage{rotating}



\usepackage{xfrac}
\usepackage[olditem,oldenum]{paralist}
\usepackage[capitalize]{cleveref}

\begin{document}

\title{Inclusive $\Lzero$ production in proton-proton collisions at 3.5 GeV}
\author{J.~Adamczewski-Musch$^{4}$, G.~Agakishiev$^{7}$, O.~Arnold$^{9,10}$,
E.T.~Atomssa$^{15}$, C.~Behnke$^{8}$, J.C.~Berger-Chen$^{9,10}$,
J.~Biernat$^{3}$, A.~Blanco$^{2}$, C.~~Blume$^{8}$, M.~B\"{o}hmer$^{10}$
% ...
}

\affiliation{
(HADES collaboration) \\\mbox{$^{1}$Institute of Physics, Slovak Academy of
Sciences, 84228~Bratislava, Slovakia}\\
\mbox{$^{2}$LIP-Laborat\'{o}rio de Instrumenta\c{c}\~{a}o e F\'{\i}sica
Experimental de Part\'{\i}culas , 3004-516~Coimbra, Portugal}\\
\mbox{$^{3}$Smoluchowski Institute of Physics, Jagiellonian University of
Cracow, 30-059~Krak\'{o}w, Poland}\\
\mbox{$^{4}$GSI Helmholtzzentrum f\"{u}r Schwerionenforschung GmbH,
64291~Darmstadt, Germany}\\
\mbox{$^{5}$Technische Universit\"{a}t Darmstadt, 64289~Darmstadt, Germany}\\
\mbox{$^{6}$Institut f\"{u}r Strahlenphysik, Helmholtz-Zentrum
Dresden-Rossendorf, 01314~Dresden, Germany}\\
\mbox{$^{7}$Joint Institute for Nuclear Research, 141980~Dubna, Russia}\\
\mbox{$^{8}$Institut f\"{u}r Kernphysik, Goethe-Universit\"{a}t,
60438 ~Frankfurt, Germany}\\
\mbox{$^{9}$Excellence Cluster 'Origin and Structure of the Universe',
85748~Garching, Germany}\\
\mbox{$^{10}$Physik Department E12, Technische Universit\"{a}t M\"{u}nchen,
85748~Garching, Germany}\\
\mbox{$^{11}$II.Physikalisches Institut, Justus Liebig Universit\"{a}t Giessen,
35392~Giessen, Germany}\\
\mbox{$^{12}$Institute for Nuclear Research, Russian Academy of Sciences,
117312~Moscow, Russia}\\
\mbox{$^{13}$Institute for Theoretical and Experimental Physics, 117218~Moscow,
Russia}\\
\mbox{$^{14}$Department of Physics, University of Cyprus, 1678~Nicosia, Cyprus}
\\
\mbox{$^{15}$Institut de Physique Nucl\'{e}aire (UMR 8608),
CNRS/IN2P3 - Universit\'{e} Paris Sud, F-91406~Orsay Cedex, France}\\
\mbox{$^{16}$Nuclear Physics Institute, Czech Academy of Sciences, 25068~Rez,
Czech Republic}\\
\mbox{$^{17}$LabCAF. F. F\'{\i}sica, Univ. de Santiago de Compostela,
15706~Santiago de Compostela, Spain}\\
\\
\mbox{$^{a}$ also at ISEC Coimbra, ~Coimbra, Portugal}\\
\mbox{$^{b}$ also at ExtreMe Matter Institute EMMI, 64291~Darmstadt, Germany}\\
\mbox{$^{c}$ also at Technische Universit\"{a}t Dresden, 01062~Dresden,
Germany}\\
\mbox{$^{d}$ also at Frederick University, 1036~Nicosia, Cyprus}\\
\mbox{$^{e}$ also at Dipartimento di Fisica and INFN, Universit\`{a} di Torino,
10125~Torino, Italy}\\
\\
\mbox{$^{\ast}$ corresponding authors: laura.fabbietti@ph.tum.de, 
rafal.lalik@ph.tum.de}
} 

\begin{abstract}
The inclusive production of $\Lzero$ hyperons in proton-proton collisions at
$\sqrt{s}=\SI{3.18}{\gev}$ was measured with \hades at the GSI
Helmholtzzentrum f\"ur Schwerionenforschung in Darmstadt. The experimental data
are compared to a data-based model for individual exclusive $\Lzero$ production
channels in the same reaction. The contributions of intermediate resonances
such as $\Sstar$, $\Dpp$ or $\Nstar$ are considered in detail. In particular,
the result of a partial wave analysis is accounted for the abundant $\pKL$
final state. Model and data show a reasonable agreement at mid rapidities,
while a difference is found for larger rapidities. A total $\Lzero$ production
\xsec in p+p collisions at $\sqrt{s}=\SI{3.18}{\gev}$ of
$\sigma(\pp\to\Lzero+X) = \num{207.3\pm1.3}\eras{+6.0}{-7.3} ~\mathrm{(stat.)}
\pm \num{8.4} ~\mathrm{(syst.)} \eras{+0.4}{-0.5}
~\mathrm{(model)}~\si{\micro\barn}$ is found.
\end{abstract}


\maketitle


\begin{figure}[!b]
 \centering
 \includegraphics[width=1.0\linewidth]{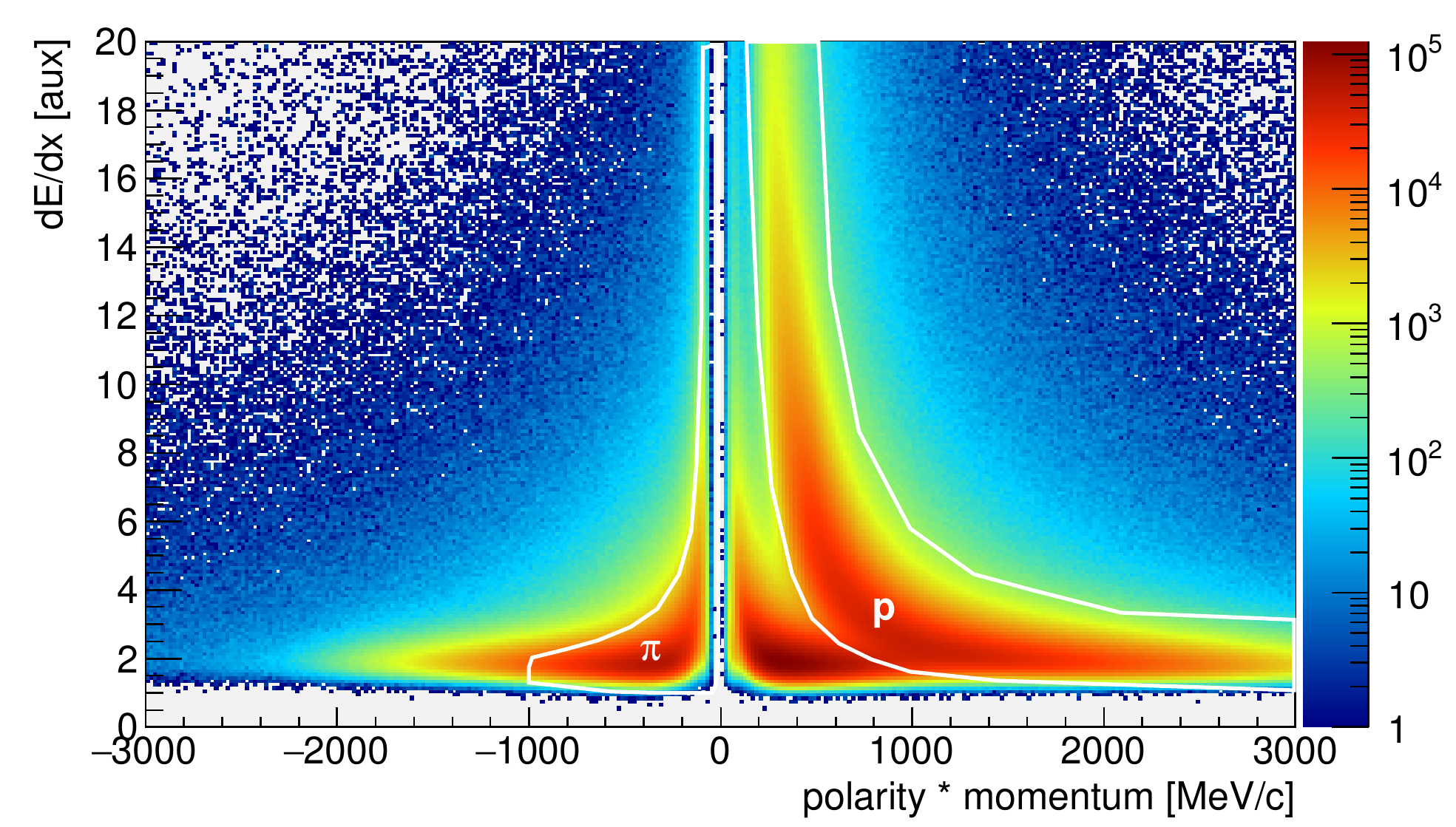}
 \caption{(Color online). Energy loss \dEdx as a function of the momentum
    multiplied by the charge for all particle candidates. The two-dimensional
    graphical cuts (white curves) are used for the particle identification.}
 \label{exp:dEdxmdc}
\end{figure}
The study of strange hadrons produced in nucleon-nucleon collisions in the
few-\si{Gev} energy range provides information about their production
mechanisms \cite{Hartnack:2011cn}. The understanding of the interaction between
strange hadrons and nucleons at different densities can have consequences also
for the modelling of the interior of neutron stars since several scenarios
include the possible presence of hyperons and kaons within the dense core
\cite{Lonardoni:2014bwa,SchaffnerBielich:2000wj,Schulze:2006vw,
Weissenborn:2011kb}. Experiments at beam energies of a few \si{\gev} are
particularly suited for these studies; while elementary reactions provide a
rather clean environment \cite{Adamczewski-Musch:2016jlh}, heavy-ion collisions
allow for a sizeable compression of nuclear matter (up to $3\rho_0$)
\cite{Li:1999bea} and hence allow to probe dense baryonic matter.
Prior to an interpretation of data from heavy-ion collisions, where more exotic
production mechanisms could occur, one should study strange hadrons, among
others $\Lzero$s, in nucleon-nucleon collisions. 
In particular, the energy regime of a few \si{\gev} is characterised by the
appearance of intermediate baryon resonances that compete with the non-resonant
production of many final states \cite{Fabbietti:2015tpa}.
A finite nuclear density larger than $\rho_0$ and a sizable temperature of the
system might modify the properties of the hadrons and also their production
mechanisms \cite{Agakishiev:2014kdy} so that precise references data are
necessary to quantify the expected in-medium effects.\\
Transport models \cite{Bass:1998ca,Buss:2011mx} are often used to
interpret the measurement of nucleus-nucleus collisions in the few-\si{\gev}
energy range and these need experimentally constrained differential \xsecs for
the different reaction channels.
Most of the models treat heavy-ion collisions as a superposition of individual
nucleon-nucleon reactions, as far as the production of secondary particles is
concerned. The question can be asked whether this approximation is appropriate 
or more complex correlations and interferences are built. A detailed comparison
of the differential spectra of strange hadrons produced in p+p and A+A
collisions can certainly help in resolving this issue
\cite{Hartnack:2011cn,Bastid:2007jz}.\\
The \hades collaboration has already carried out several exclusive measurements
of final states containing strange hadrons. These studies in p+p collisions at
a kinetic beam energy of \SI{3.5}{\gev} have so far focused on the exclusive
production of $\Kzs$ \cite{PhysRevC.90.015202,Agakishiev:2014moo}, $\Kstar$
\cite{Agakishiev:2015ysr}, $\Lzero$ \cite{Agakishiev2015242}, $\Sstarp$
\cite{PhysRevC.85.035203,PhysRevC.87.025201,PhysRevC.90.015202} and $\Lstar$
\cite{PhysRevC.85.035203,Siebenson:2013rpa}, where for some of these channels
also measurements of the angular distributions were possible. 
One of the goals of these works was to study the contribution of intermediate
resonances coupling to the different final states, and in certain cases a clear
signature of the impact of these resonances was found
\cite{PhysRevC.85.035203,Agakishiev:2014dha}.
As for $\Lzero$ hyperons, a partial wave analysis (PWA) was employed to analyse
the exclusive reaction $\pp\to\pKL$ and the contribution of the resonances
$\Nstar(1650,1710,1720,1850,1900,1950)$ was evaluated.
The important contribution of $\Nstar$ resonances to the $\pKL$ final state was
already pointed out in previous analyses
\cite{AbdelBary:2010pc,Fabbietti2010333} but the PWA allows for a more
quantitative determination of the contributing resonant and non-resonant
channels.
In addition to the $\pp \to \pKL$ exclusive reaction, many other channels
contribute to the inclusive $\Lzero$ production in $\prot+\prot$ collisions. In
the present work, we propose a model for inclusive $\Lzero$ production which
combines the information extracted from the PWA of the $\pKL$ final state
\cite{AbdelBary:2010pc,Fabbietti2010333} with the measurements of other
exclusive channels containing a $\Lzero$ \cite{Agakishiev2015242,
AbdelBary:2010pc,PhysRevC.85.035203,PhysRevC.90.015202,PhysRevC.87.025201,
PhysRevC.90.054906} and uses estimates for unmeasured channels.
All channels have been simulated independently; the $\pKL$ yield is obtained
from PWA while all other channels are added in an incoherent way. This
cocktail is then used to fit the experimental data and to extract the \xsec for
the $\Lzero$ production and for its various contributing channels.\\
Our paper is organized as follows. In \cref{sec1} the experimental data are
described. \Cref{sec2} presents the production model used for the comparison to
the experimental data and the evaluation of the efficiency and acceptance
corrections. The latter ones are described in \cref{sec3}. In sections
\cref{sec4,sec5} the comparison of the model to the experimental data and the
extraction of the inclusive $\Lzero$ production \xsec in p+p collisions at
\SI{3.5}{\gev} are discussed.

\section{Experimental data}
\label{sec1}
\begin{figure*}[hbt]
 \centering
 \includegraphics[width=1.0\linewidth]{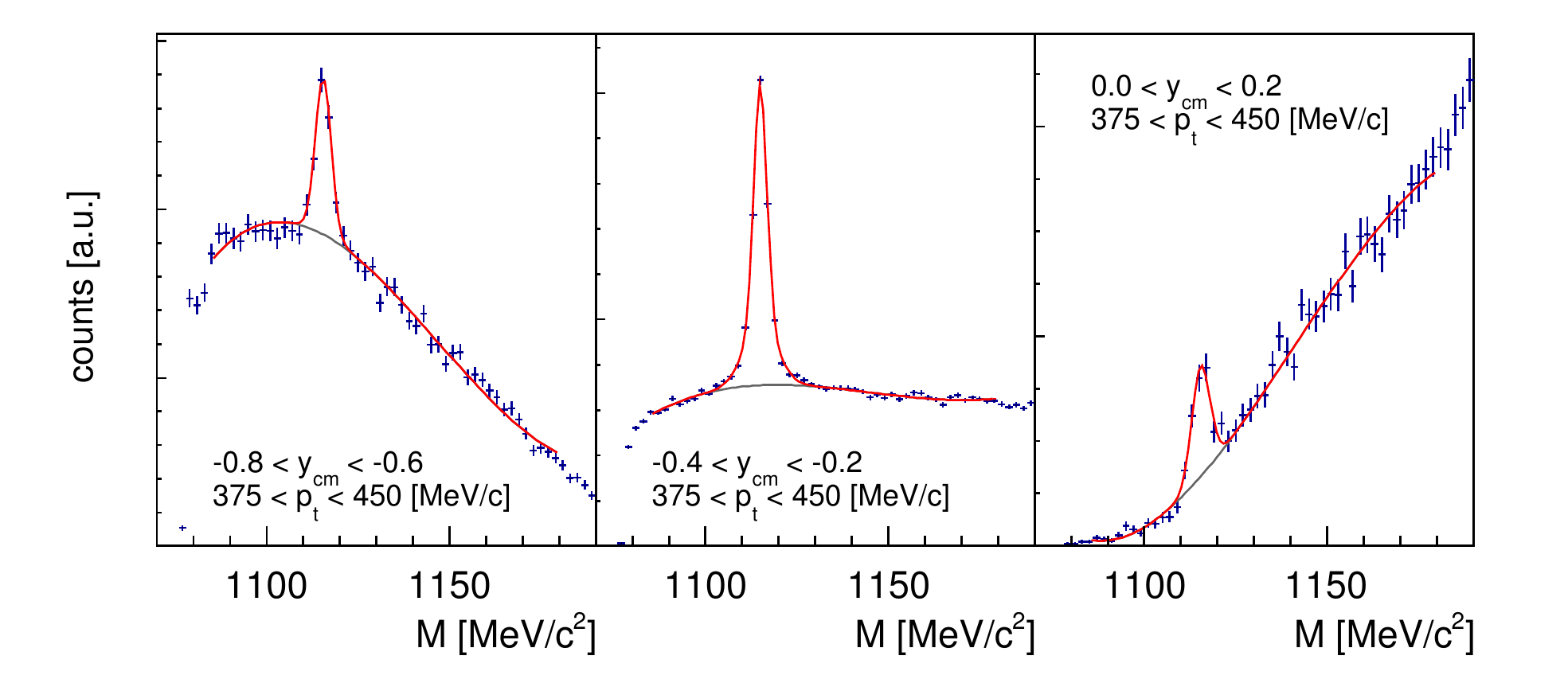}
 \caption{(Color online). Invariant mass spectrum for proton-pion pairs for
    three different \ycm bins. The red solid line shows the signal+background
 fit, the black dashed line shows the background component of the fit.}
 \label{exp:data_yield}
\end{figure*}

The experimental data were obtained with the {\bf H}igh {\bf A}cceptance
{\bf D}i-{\bf E}lectron {\bf S}pectrometer (\hades) at the heavy-ion 
synchrotron SIS18 at GSI Helmholtzzentrum f\"ur Schwerionenforschung in
Darmstadt, Germany.
\hades is a charged-particle detector consisting of a six identical detection
sections (with a nearly complete azimuthal coverage) centred on the beam axis
and covering polar angles between \SI{18}{\degree} and \SI{85}{\degree}, and a
six-coil toroidal magnet located between two pairs of tracking chambers. Each
sector is equipped with a Ring-Imaging Cherenkov (RICH) detector followed by
Multi-wire Drift Chambers (MDCs) -- two in front of and two behind the magnetic
field -- as well as the two scintillator hodoscopes TOF and TOFino and a
PreShower detector. The hadron identification is based on the correlation
between momentum and specific energy-loss information obtained from the MDC
tracking detectors. In the following, the TOF-TOFino-PreShower system is
referred to as Multiplicity Electron Trigger Array (META). A detailed
description of \hades can be found in \cite{Agakishiev:2009am}. 

During an experimental campaign in 2007, a proton beam of about \num{e6}
particles/s with \SI{3.5}{GeV} kinetic energy was incident on a liquid hydrogen
target of \SI{50}{mm} thickness corresponding to \SI{0.7}{\percent} interaction
probability.
The data readout was started by a first-level trigger (LVL1) requiring a
charged-particle multiplicity, $\mathrm{MUL} \ge 3$, in the META system. A
total of \num{1.14e9} events were recorded under these experimental conditions.
A dedicated calibration run without target was also carried out and \num{1.2e5}
events were analysed to study the contribution by off-target reactions.
$\Lzero$ hyperons were reconstructed exploiting the decay
$\Lzero\to\prot\pim$, $BR =\SI{63.9}{\percent}$ \cite{Agashe:2014kda}.
Since $c\tau_\Lzero = \SI{78.9}{\milli\metre}$, it was possible to apply
topological cuts to reduce the background contribution.
The first step of the analysis consisted in the selection of the proton and 
negative pion candidates. This was done applying graphical cuts on the energy
loss distribution measured in the MDC (see white curves in \cref{exp:dEdxmdc})
as a function of the particle momentum.
\begin{figure*}[!t]
 \centering
 \begin{subfigure}[b]{0.495\textwidth}
  \includegraphics[width=1.0\textwidth]{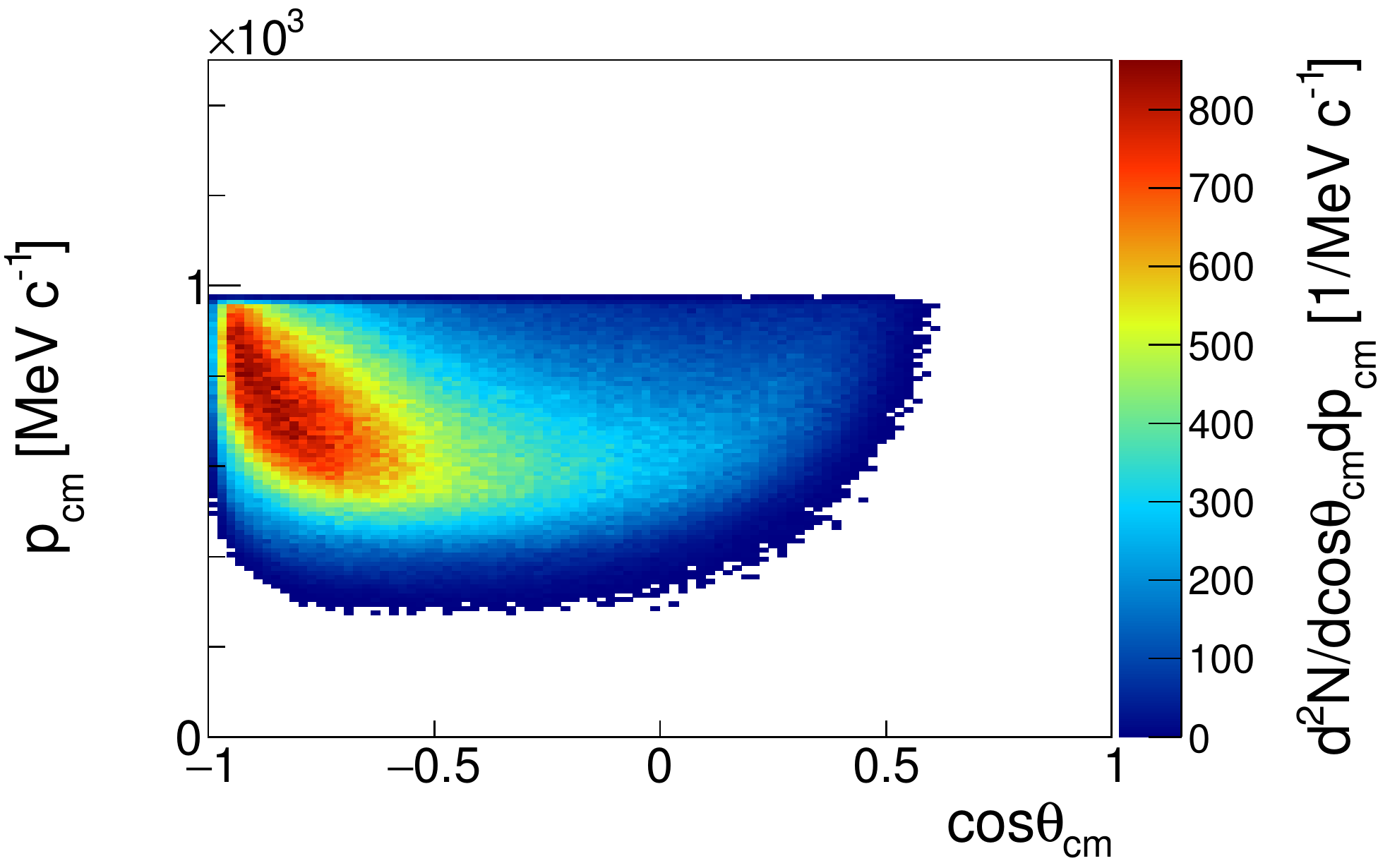}
 \label{fig:ps_diff:pty}
 \end{subfigure}
%
%
 \begin{subfigure}[b]{0.495\textwidth}
  \includegraphics[width=1.0\textwidth]{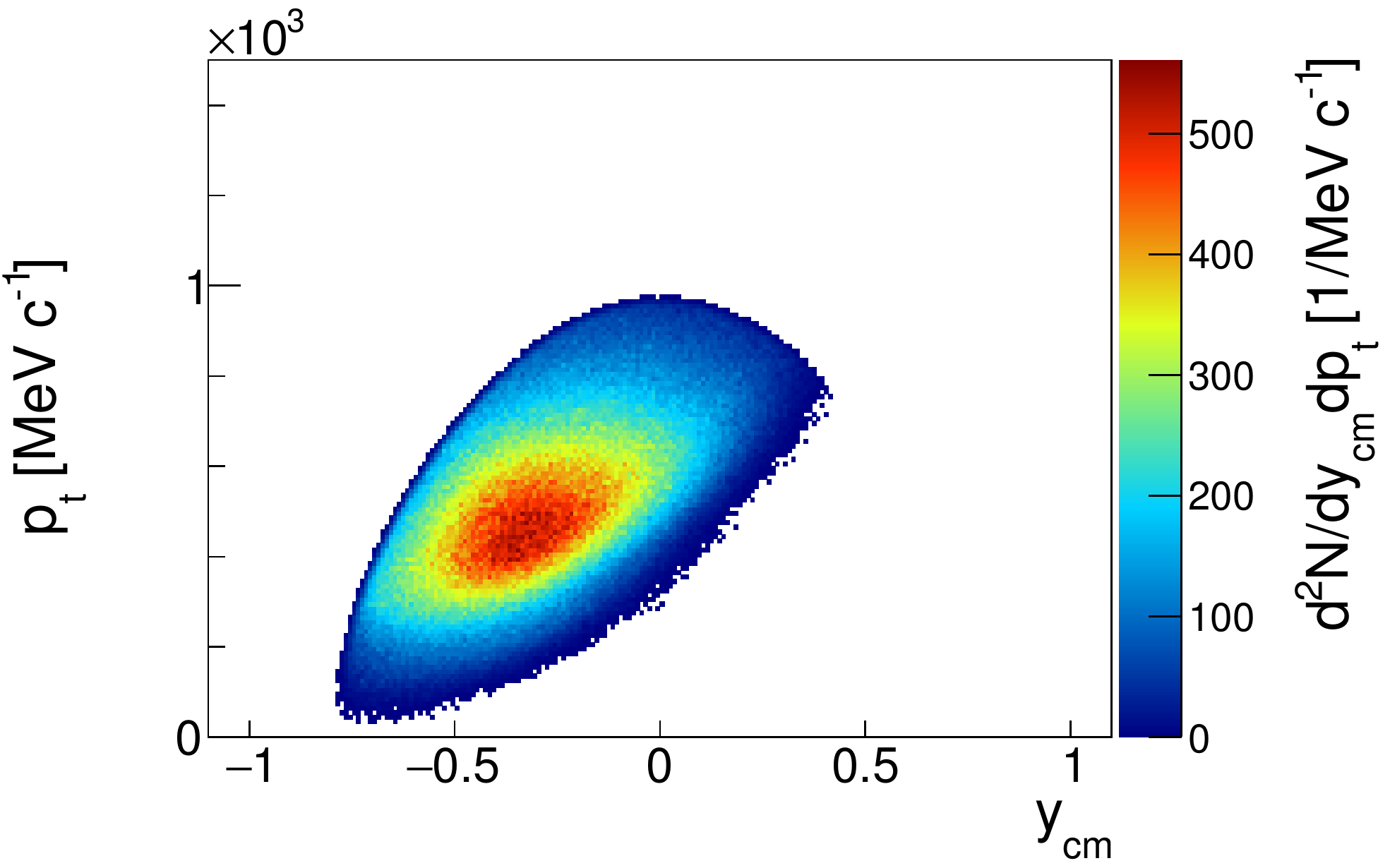}
 \label{fig:ps_diff:pcosth}
 \end{subfigure}

 \caption{(Color online). Phase space distributions of $\Lzero$ candidates prior
    to efficiency corrections. Left panel: $\Lzero$ momentum in the p-p cms as a
    function of the polar angle \cthcm. Right panel: transverse $\Lzero$
    momentum as a function of the cms rapidity \ycm.}
 \label{fig:ps_diff}
\end{figure*}
The $\Lzero$ decay point (called secondary vertex (SV)) was associated to the
point of closest approach between the proton and pion tracks. The
reconstructed $\Lzero$ track was then extrapolated to the target
region and the intersection point with all other reconstructed tracks in the
same event was computed. This intersection point was associated to the primary
vertex (PV) of the \pp reaction. If no other tracks were available for a given
event, the intersection between the $\Lzero$ track candidate and the average
beam trajectory was used. The latter is recalculated for each day of data
taking. The resolution for the reconstruction of the primary vertex was
extracted by analyzing the empty-target data and taking the Kapton windows that
enclose the \ce{LH_2} target as a reference. A resolution of \SI{3.8}{\mm} in
the $z$-direction was found. The resolution in the beam direction of the
$\Lzero$ secondary vertex reconstruction was evaluated with simulations and a
value of \SI{1.8}{\mm} was determined.
\\
A further analysis of the empty-target events showed the presence of a
contamination stemming from hyperons produced in the \SI{50}{\micro\metre} thick
Kapton windows (weight fractions: \SI{69}{\percent} of \ce{^{12}C},
\SI{21}{\percent} of \ce{^{16}O}, \SI{7}{\percent} of \ce{^{14}N} and
\SI{3}{\percent} of \ce{^{1}H}) at both ends of the \ce{LH_2} target.
The total contribution of these processes was estimated, considering the
relative thickness and density of the targets and assuming a scaling of the
\xsec with the atomic number A, to be between \SI{3}{\percent} and
\SI{6}{\percent} \cite{PoS_Bormio2015_009}.\\
The dimensions of the \ce{LH_2} target extended within
$\SI{-65}{mm}<z<\SI{-15}{mm}$ in length and $r<\SI{5}{mm}$ in radius.
Only events with a reconstructed primary vertex localized within the
volume defined as a tube of $\SI{-50}{mm}<z<\SI{-30}{mm}$ length and with a
radius of $r<\SI{5}{mm}$, were considered further in the analysis to minimize
the contribution of the Kapton windows.
This selection of the z-coordinate reduces the off-vertex contribution from the
Kapton windows to a level of less than one per mille.\\
In order to minimize the combinatorial background emerging from misidentified or
uncorrelated $\prot\pim$ pairs, three topological cuts were
applied:%
\begin{inparaenum}[(\itshape 1)]
 \item the $z$-coordinate of the decay vertex must be larger than the
    $z$-coordinate of the primary vertex. This cut reduces the background by a
    factor of two and affects the signal only by \SI{2}{\percent},
 \item the distance of closest approach (DCA) between the $\prot$ and $\pim$
    tracks should be smaller than \SI{10}{\milli\metre},
 \item the pointing angle (PA) between the spatial vector connecting PV
    and SV and the momentum vector of $\Lzero$ should be smaller than
    \SI{0.1}{\radian}.
\end{inparaenum}
Additionally, the missing mass (MM) of the reaction $\pp\to\Lzero+X$ should be
larger than \SI{1400}{\mevsc}.

\Cref{exp:data_yield} shows the resulting $\prot\pim$ invariant mass
distribution for three bins in the \pp \tcms rapidity together with the fit used
to extract the signal strength. 
The $\Lzero$ peak was fitted with the weighted sum of two Gaussian
distributions $G(x, \mu, \sigma)$ with a common mean value $\mu$ and two
different width parameters $\sigma_1$ and $\sigma_2$ and a relative
contribution $c$:
\begin{equation}
 S = N \left[c \cdot G(x, \mu, \sigma_1) + (1-c) \cdot G(x, \mu, \sigma_2) \right]\text{,}
 \label{eq:signalfit}
\end{equation}
where $N$ is the total amplitude of the signal. The background distribution was
modelled using the sum of a polynomial of fifth order and an exponential
function.
The signal yield and the error were extracted from the fit parameters and their
errors. The signal to background ratio was calculated integrating the fitted
signal and the background functions over a $ \mu \pm 3 \sigma$ range, where
$\sigma = c\sigma_1 + (1-c)\sigma_2$.

For the integrated $\prot\pim$ invariant mass distribution the mean value $\mu$
of the reconstructed $\Lzero$ mass is \SI{1115.122 \pm 0.009}{\mevsc} and the
parameters $\sigma_1$, $\sigma_2$ and $c$ are found to be equal to
\SI{4.075\pm 0.094}{\mevsc}, \SI{1.665 \pm 0.023}{\mevsc} and 
\num{0.487 \pm 0.015}, respectively. The reconstructed $\Lzero$ mass is in good
agreement with the PDG value \cite{Agashe:2014kda}.
The total yield of reconstructed $\Lzero$s amounts to \num{258.2 \pm 1.2 e3} 
with a signal to background ratio of \num{0.47}.\\
Thanks to the large statistics, two differential analyses of the $\Lzero$ yield
were carried out :%
\begin{inparaenum}[(\itshape 1)]
 \item as a function of the particle momentum and cosine of the polar angle, and
 \item as a function of the transverse momentum and the rapidity.
\end{inparaenum} 
All the kinematic variables were calculated in the beam+target (p+p) \tcms
system (cms). The resulting \phasespace distributions are shown in
\cref{fig:ps_diff} after applying an additional cut on the $\Lzero$ invariant
mass $\SI{1102}{\mevsc} < M_{\prot\pim} < \SI{1130}{\mevsc}$. These
distributions are not corrected for acceptance and efficiency.
The geometrical acceptance of \hades allows the reconstruction of
$\Lzero$ hyperons with a momentum between \SI{300}{\mevc} and \SI{930}{\mevc} 
measured in the center of mass system of the p+p collision system.
The polar coverage of \SIrange{18}{85}{\degree} in the laboratory system
translates into \SIrange{53}{180}{\degree} in the proton-proton cms (i.e.
cosine between \numrange{-1}{0.6}).\\
In order to obtain differential distributions of the $\Lzero$ hyperon signal,
the \phasespace{} distributions were divided into discrete bins of equal size.
For the \cthcm and \pcm variables, the intervals between $[-1.0,0.6]$ and
$[0,1.0]$ \si{\gevc} were divided into \num{8} and \num{10} bins, respectively.
For the \ycm and \ptcm variables, the intervals between $[-0.8,0.4]$ and
$[0,1.0]$ \si{\gevc} were divided into \num{6} and \num{13} bins, respectively. 
A $\prot\pim$ invariant mass spectrum was obtained for each of the selected bins
by applying the topological selections mentioned above and the same fit
procedure as for the integrated spectrum.
The signal and the corresponding errors were also calculated in the same way.
These values stay consistent within a few percent with respect to the total
invariant mass spectrum fit.

\section{Production model} \label{sec2}

{
\renewcommand{\arraystretch}{1.2}%

\squeezetable

\begin{table*}[!t]
 \caption{List of channels included in the production model for $\Lzero$s in p+p
    collisions at $\sqrt{s}=\SI{3.18}{\gev}$. The total \xsec and asymmetry
    parameters $a_{2,4}$ are listed for each channel. The column labelled by H
    denotes channels exclusively measured by \hades. The "notes" column shows
    source references, for the other comments please refer to the text for
    details. The last column lists the cross section resulting from a model fit
    to the data.}
 \label{tab:model}

 \begin{tabularx}{1.0\textwidth}{p{2em}lXp{5em}p{6em}p{6em}cc||p{8em}c}
  id & \prot\prot $\to$ reaction & $\sigma_0^{(\mathrm{id})}$ \xsec
    $[\si{\micro\barn}]$ &
  $\measuredangle$~var. & \multicolumn{2}{c}{$\measuredangle (a_2, a_4)$} & H &
  notes & \multicolumn{2}{c}{fit result} \\
  \toprule
  \multicolumn{8}{c}{3-body channels} \\
  \midrule
  1 & $\Lzero\prot\Kp$    & $35.26 \pm 0.43 \eras{+3.55}{-2.83}$
    & $\theta_\Lzero^\cms$ & $0.798$ & $0.134$ & $\checkmark$
    & \cite{Agakishiev2015242}
    & \num{38.835 +- 0.026} & $\top$ \\
  2 & $\Szero\prot\Kp$        & $16.5  \pm \SI{20}{\percent}$
    & $\theta_\Szero^\cms$ & $0.034 \pm 0.241$ & --- &
    & \cite{AbdelBary:2010pc}+calc.
    & \num{19.800 +- 0.094} & $\top$ \\
  \midrule
  3 & $\Lzero\Dpp\Kz$        & $29.45 \pm 0.08 \eras{+1.67}{-1.46} \pm 2.06$
    & $\theta_\Dpp^\cms$ & $1.49 \pm 0.3$ & --- & $\checkmark$
    & \cite{PhysRevC.90.015202}
    & \num{32.10 +- 0.11} & $\top$ \\
  4 & $\Szero\Dpp\Kz$ & $\phantom{0}9.26 \pm 0.05 \eras{+1.41}{-0.31} \pm 0.65$
    & $\theta_\Dpp^\cms$ & $0.08 \pm 0.02$ & --- & $\checkmark$
    & \cite{PhysRevC.90.015202}
    & $\phantom{0}\num{8.5 +- 2.1}$ & $\bot$ \\
  5 & $\Lzero\Dp\Kp$        & $\phantom{0}9.82 \pm \SI{20}{\percent}$
    & $\theta_\Dp^\cms$ & \multicolumn{2}{c}{from $\Lzero\Dpp\Kz$} &
    & res. mod.
    & \num{11.78 +- 0.15} & $\top$ \\
  6 & $\Szero\Dp\Kp$        & $\phantom{0}3.27 \pm \SI{20}{\percent}$
    & $\theta_\Dp^\cms$ & \multicolumn{2}{c}{from $\Szero\Dpp\Kz$} &
    & res. mod.
    & $\phantom{0}\num{2.6 +- 1.3}$ & $\bot$ \\
  \midrule
  7 & $\Sstarp\neut\Kp$    & $22.42 \pm 0.99 \pm 1.57 \eras{+3.04}{-2.23}$
    & $\theta_{\Splus^*}^\cms$ & $1.427 \pm 0.3$ & $0.407 \pm 0.108$
    & $\checkmark$ & \cite{PhysRevC.85.035203}
    & \num{17.905 +- 0.075} & $\bot$ \\
  8 & $\Delta(2050)^{++}\neut$
    & \SI{33}{\percent} feeding for $\Sigma^*\neut\Kp$
    & $\theta_{\neut}^\cms$ & $1.27$ & $0.35$ & $\checkmark$
    & \cite{PhysRevC.85.035203}
    & $\phantom{0}\num{8.82 +- 0.13}$ & $\top$ \\
  9 & $\Sstarp\prot\Kz$    & $14.05 \pm 0.05 \eras{+1.79}{-2.14} \pm 1.00$
    & $\theta_{\Splus^*}^\cms$ & $1.42 \pm 0.3$ & --- & $\checkmark$
    & \cite{PhysRevC.90.015202}
    & \num{16.101 +- 0.072} & $\top$ \\
  10 & $\Sstarz\prot\Kp$    & $\phantom{0}6.0 \pm 0.48 \eras{+1.94}{-1.06}$
    & $\theta_{\Szero^*}^\cms$
    & \multicolumn{2}{c}{from $\Sstarp\neut\Kp$} & $\checkmark$
    & \cite{PhysRevC.85.035203}
    & $\phantom{0}\num{7.998 +- 0.069}$ & $\top$ \\
  11 & $\Lstar\prot\Kp$     & $\phantom{0}9.2 \pm 0.9 \pm 0.7 \eras{+3.3}{-1.0}$
    & --- &  ---  &  --- & $\checkmark$
    & \cite{PhysRevC.87.025201}
    & $\phantom{0}\num{7.7 +- 3.0}$ & $\bot$ \\
  12 & $\Lstard\prot\Kp$    & $\phantom{0}5.6 \pm 1.1\pm0.4\eras{+1.1}{-1.6}$
    & --- &  ---  &  --- & $\checkmark$
    & \cite{PhysRevC.87.025201}
    & $\phantom{0}\num{7.2 +- 3.6}$ & $\top$ \\
  \midrule
  13 & $\Dpp\Lstar\Kz$        & $\phantom{0}5.0 \pm \SI{20}{\percent}$
    & --- &  ---  &  --- &
    & \cite{PhysRevC.90.054906}
    & $\phantom{0}\num{6.0 +- 1.6}$ & $\top$ \\
  14 & $\Dpp\Sstarz\Kz$        & $\phantom{0}3.5 \pm \SI{20}{\percent}$
    & --- &  ---  &  --- &
    & \cite{PhysRevC.90.054906}
    & $\phantom{0}\num{4.90 +- 0.46}$ & $\top$ \\
  15 & $\Dp\Sstarp\Kz$        & $\phantom{0}2.3 \pm \SI{20}{\percent}$
    & --- &  ---  &  --- &
    & \cite{PhysRevC.90.054906}
    & $\phantom{0}\num{3.2 +- 1.1}$ & $\top$ \\
  16 & $\Dp\Lstar\Kp$        & $\phantom{0}3.0 \pm \SI{20}{\percent}$
    & --- &  ---  &  --- &
    & compl. to above
    & $\phantom{0}\num{4.2 +- 1.9}$ & $\top$ \\
  17 & $\Dp\Sstarz\Kp$        & $\phantom{0}2.3 \pm \SI{20}{\percent}$
    & --- &  ---  &  --- &
    & compl. to above
    & $\phantom{0}\num{3.2 +- 1.1}$ & $\top$ \\
  \toprule
  \multicolumn{8}{c}{4-body channels} \\
  \midrule
  18 & $\Lzero\prot\pip\Kz$
    & $\phantom{0}2.57 \pm 0.02 \eras{+0.21}{-1.98} \pm 0.18$
    & \multicolumn{3}{c}{---} & $\checkmark$
    & \cite{PhysRevC.90.015202}
    & $\phantom{0}\num{2.8 +- 1.5}$ & $\top$ \\
  19 & $\Lzero\neut\pip\Kp$    & from $\Lzero\prot\pip\Kz$
    & \multicolumn{3}{c}{---} & &
    & $\phantom{0}\num{2.8 +- 1.5}$ & $\top$ \\
  20 & $\Lzero\prot\piz\Kp$    & from $\Lzero\prot\pip\Kz$
    & \multicolumn{3}{c}{---} & &
    & $\phantom{0}\num{2.8 +- 1.4}$ & $\top$ \\
  21 & $\Szero\prot\pip\Kz$
    & $\phantom{0}1.35 \pm 0.02 \eras{+0.10}{-1.35} \pm 0.09$
    & \multicolumn{3}{c}{---} & $\checkmark$
    & \cite{PhysRevC.90.015202}
    & $\phantom{0}\num{1.48 +- 0.76}$ & $\top$ \\
  22 & $\Szero\neut\pip\Kp$    & from $\Szero\prot\pip\Kz$
    & \multicolumn{3}{c}{---} & &
    & $\phantom{0}\num{1.48 +- 0.84}$ & $\top$ \\
  23 & $\Szero\prot\piz\Kp$    & from $\Szero\prot\pip\Kz$
    & \multicolumn{3}{c}{---} & &
    & $\phantom{0}\num{1.48 +- 0.75}$ & $\top$ \\
  \bottomrule
 \end{tabularx}
\end{table*}
}

A model for $\Lzero$ production in p+p collisions at $\sqrt{s}=\SI{3.18}{\gev}$
was built based partially on exclusive measurements carried out by \hades
\cite{Agakishiev2015242,PhysRevC.90.015202,PhysRevC.85.035203,
PhysRevC.87.025201,PhysRevC.90.054906}, and partially relying on a resonance
model \cite{Tsushima:1996xc,PhysRevC.59.369} and results by the COSY
collaboration \cite{AbdelBary:2010pc}.
The various contributions in the model can be divided into five categories:%
\begin{inparaenum}[\itshape a\upshape)]
 \item three-body direct production ($\mathrm{pKY}$),
 \item associated resonance production ($\Delta$KY),
 \item intermediate resonance production (pKY$^*$),
 \item double resonance production ($\Delta$KY$^*$),
 \item 4-, 5-body, and higher order \phasespace production.
\end{inparaenum}
%
%
Each channel is characterized by a total production \xsec
$\sigma_0^{(\mathrm{id})}$ and the coefficients $a_0$, $a_2^{(\mathrm{id})}$,
$a_4^{(\mathrm{id})}$ that are associated to Legendre polynomials $P_i(x)$ to
describe the anisotropy of the angular distributions. The employed
parametrisation is:
\begin{equation}
\sigma(x)^{(\mathrm{id})} = \frac{\sigma_0^{(\mathrm{id})}}{2} \left\{
    a_0 P_0(x) + a_2^{(\mathrm{id})} P_2(x) + a_4^{(\mathrm{id})} P_4(x)
    \right\} \text{,}
 \label{eq:csang}
\end{equation}
where $x \equiv \cos(\theta^\cms)$. The $\sfrac{1}{2}$ factor is for
normalisation of the shape component and $a_0$ is fixed to \num{1}.
In the proton-proton cms a symmetric angular distribution with respect to $x=0$
must hold true and, therefore, odd-order polynomials were ignored and only the
three first even terms were used.\\
\Cref{tab:model} shows a complete list of the different production channels
included in the model. The channels with a tag in the column H are those
measured exclusively in the same data sample by \hades
\cite{Agakishiev2015242,PhysRevC.90.015202,PhysRevC.85.035203,
PhysRevC.87.025201,PhysRevC.90.054906}.
For reactions which are not measured at \hades energies, either theoretical
predictions from a resonance model \cite{Tsushima:1996xc,PhysRevC.59.369} are
used or isospin symmetries are exploited to estimate the \xsecs.\\
An exclusive measurement of the channel $\pp\to\pKL$ allowed to study the
contribution of different $\Nstar$ resonances to this final state
\cite{Agakishiev2015242}; the results differ strongly from a phase space 
model \cite{Fabbietti:2013npl}. In
our analysis the solutions of a PWA fit were used in the model for the channel
$\pp\to\pKL$. The \xsec for this channel was evaluated in
\cite{Agakishiev2015242}.
The parameters for the angular distribution shown in \cref{tab:model} were
obtained by fitting the $\Lzero$ polar angle distribution obtained from the PWA.
The reaction $\pp\to\Szero\prot\Kp$ (id=2) was measured by the COSY-TOF
experiment at $\sqrt{s} \approx \SI{2.7}{\gev}$ \cite{AbdelBary:2010pc}. These
data and the other measurements listed in \cite{lb_v2_rea_1_32} show an average
ratio of about $\num{2.2}$ between the $\Lzero$ and $\Szero$ \xsecs.
For that reason, the \xsec for channel id=2 was evaluated by dividing the \xsec
for channel id=1 by a factor \num{2.2} and assuming an uncertainty of
\SI{20}{\percent}. For the angular distribution, the anisotropy parameters
measured by the COSY collaboration \cite{AbdelBary:2010pc} were used for this
channel.

Channels with an associated $\Dpp$ resonance production (id=3--4) were measured
exclusively by \hades and also the angular distributions were extracted 
\cite{PhysRevC.90.015202}. No data are available for the associated production
with a $\Dp$ resonance (id=5--6), hence these \xsecs were constrained to the
known \xsec for the respective $\Dpp$ channels applying the following isospin
relations: $\sigma(\pp\to\hyp\Dpp\Kz)/\sigma(\pp\to\hyp\Dp\Kp)=\num{3}$. An
error of \SI{20}{\percent} was assumed. The angular coefficients for channels 5
and 6 were assumed to be equal to those of the corresponding $\Dpp$ channels
(id=3,4).\\
Channels with an intermediate resonance decaying into a final state containing a
$\Lzero$ hyperon are listed in \cref{tab:model} with id=7--12 and all of them
were measured exclusively by \hades \cite{PhysRevC.85.035203,
PhysRevC.87.025201,PhysRevC.90.015202}.
The angular anisotropy for the reaction $\Sstarp\neut\Kp$, followed by the decay
$\Sstarp\rightarrow \Lambda\pip$,  was measured for the $\Sstar$ and the same
distribution was used to simulate this channel.
The isospin symmetric reactions with id=7,9 and 10 were assigned the same
angular coefficients for the $\Lzero$ in the final state.
An isotropic angular distribution was assumed for the channels with an
intermediate production of a $\Lstar$ or a $\Lstard$, since the exclusive
measurement of these channels did not show any significant anisotropy
\cite{PhysRevC.87.025201}. Nevertheless, one should mention that the statistics
of the exclusive measurement of these channels was rather limited and hence the
assumption could be wrong.\\
The channels with id=13--17 represent the $\Lzero$ production via an
intermediate resonance with an associated $\Delta$ production.
The \xsecs for the channels 13--15 were extracted from a fit to 
an inclusive measurement of $\Kzs$
spectra, also performed by \hades at the same energy \cite{PhysRevC.90.054906}.
Channels with id=16 and 17 were considered to have the same yield as
channel id=15. An estimated error of \SI{20}{\percent} was assigned to all these
\xsec values and no angular distributions were implemented.
The four-body \phasespace production was analysed only in the neutral kaon
channel for $\Lzero$ and $\Szero$ \cite{PhysRevC.90.015202}. The production
threshold for the neutral kaon channel is only \SI{9}{\mev} higher than the one
for the charged kaon channels. This justifies the assumption that the
\phasespace distribution and production \xsecs should be the same for all
$\Lzero$ or $\Szero$ four-body processes. Therefore, only channels with id=18
and 21 were considered in the global model and their \xsec was scaled by a
factor of three to take into account the contribution from channels with
id=19--20 and id=22--23, respectively.\\
Other channels, with a larger number of particles in the final state, were
neglected since their total contribution is expected to be negligible at
energies  below a few \si{\gev}.
\begin{figure*}[!t]
  \centering
  \includegraphics[width=1.0\textwidth]{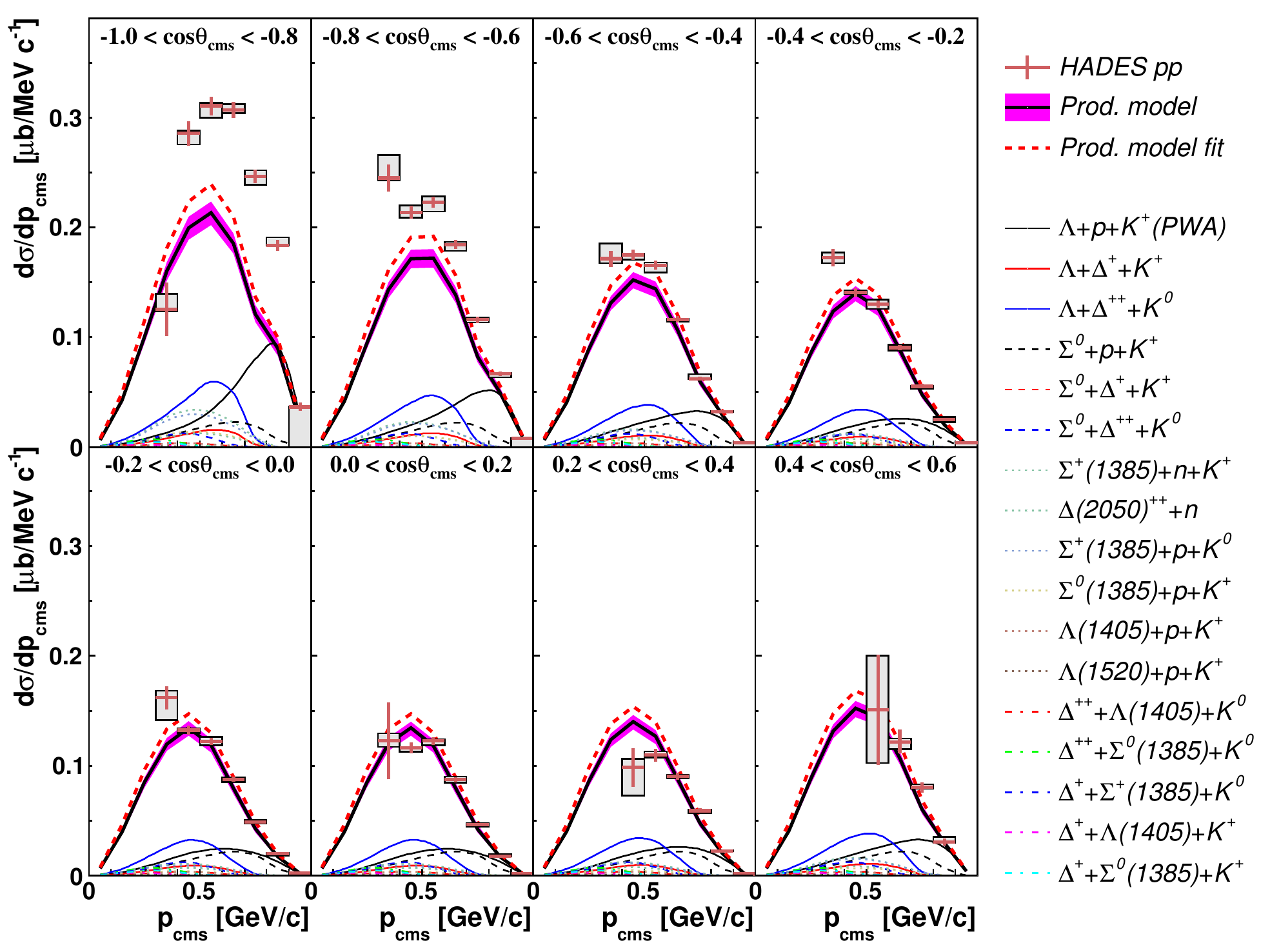}
  \caption[Corrected experimental data and production model of the differential
    \xsec $\ud\sigma/\ud$ \pcth \phasespace.]{(Color online). Corrected
    experimental data and production model of the differential \xsec
    $\ud\sigma/\ud$\pcm for different bins of \cthcm. Statistical and systematic
    errors are depicted by crosses and boxes, respectively. The solid black and
    red dashed curves represent the original and refitted production model,
    respectively. The contribution by the different channels to the original
    model are depicted by the different curves labeled in the figure legend.}
  \label{fig:comp_1}
\end{figure*}

\begin{figure*}[!t]
  \centering
  \includegraphics[width=1.0\textwidth]{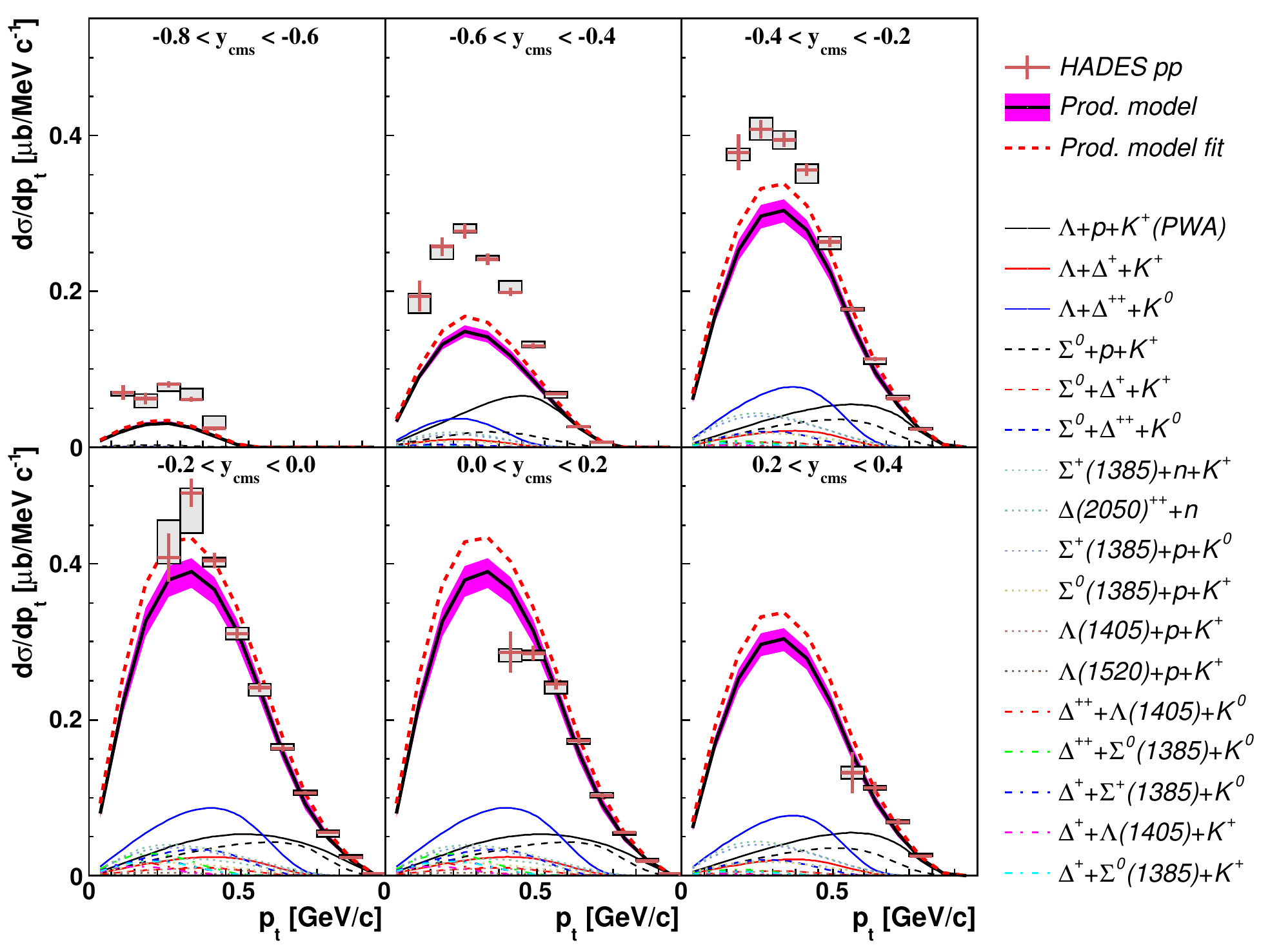}
  \caption[Corrected experimental data and production model in the \pty
    \phasespace.]{(Color online). Same as \cref{fig:comp_1} for $\ud\sigma/\ud$
    \ptcm for different bins of \ycm.}
  \label{fig:comp_2}
\end{figure*}

\section{Corrections and Error Estimation}\label{sec3}

{
\renewcommand{\arraystretch}{1.2}
\begin{table}[!b]
 \caption{Cuts and their variations for the systematical uncertanity
    evaluation.}
 \label{tab:cutsvar}

 \begin{tabularx}{1.0\linewidth}{XCCC}
  \toprule
  Cuts variant & DCA [\si{\milli\metre}] & PA [\si{\radian}]
    & MM [\si{\mevsc}] \\
  \midrule
  Regular cuts    & $<\num{10}$ & $<\num{0.10}$ & $>\num{1400}$ \\
  Loose cuts    & $<\num{12}$ & $<\num{0.12}$ & $>\num{1260}$ \\
  Strict cuts    & $<\phantom{1}\num{8}$ & $<\num{0.08}$ & $>\num{1540}$ \\
  \bottomrule
 \end{tabularx}
\end{table}
}
The production model with its input values for the \xsecs
$\sigma_0^{(\mathrm{id})}$ and angular coefficients $a_{2,4}^{(\mathrm{id})}$
as listed in \cref{tab:model} was used to determine an acceptance correction
matrix. Each channel was simulated using the Pluto event generator
\cite{Frohlich:2007bi}. First, isotropic distributions of the different final
states were simulated and in a second step weights were introduced for each
simulated event to account for the channel \xsec and the angular distribution.
For each reconstructed $\Lzero$ candidate the weight is calculated according to
\cref{eq:csang} using the $\theta$ angle depending on the specific channel as
indicated in the fourth column of \cref{tab:model}, where the $\sigma_0$
scaling factor and $a_2^{(\mathrm{id})}$ and $a_4^{(\mathrm{id})}$ coefficients
are taken also from \cref{tab:model}. The resulting events are then passed
through a full-scale simulation composed of a Geant3 \cite{Brun:118715} part
taking care of the particle interaction in the different sub-detectors and a
digitisation part that accounts for the sub-detector response. For the
simulated events the same analysis steps were followed as for the experimental
data. The resulting contributions of the different channels are summed up
according to their specific \xsecs.\\
In order to compare the experimental data to the production model, corrections
for the geometrical acceptance and reconstruction efficiency had to be applied.
The correction matrices were obtained by dividing the discrete \phasespace
distribution of the simulated data after the full scale analysis and the input
provided by the production model before the filtering. This matrix was employed
to correct the experimental data and an additional normalisation factor
extracted from elastic \pp events \cite{aipcp:10.1063:1.3483432} allowed to
evaluate the differential \xsecs for the inclusive $\Lzero$ production.
The normalization error is equal to \SI{7.28}{\percent}.\\
{
\renewcommand{\arraystretch}{1.2}

\begin{table}[!b]
 \caption{Systematic uncertainty evaluation.}
 \label{tab:systematic}

 \begin{tabularx}{1.0\linewidth}{lCC}
  \toprule
  Uncertainty source & \pcth analysis & \pty analysis \\
  \midrule

  Topological cuts variation    & $\eras{+6.1}{-7.5}$~\si{\percent} &
    $\eras{+4.3}{-5.0}$~\si{\percent} \\
    \midrule
  Acceptance matrix sampling    & \SI{1.0}{\percent} & \SI{0.9}{\percent}\\
  \midrule
  Normalisation  & \SI{7.3}{\percent} & \SI{7.3}{\percent}  \\
  
  \midrule
  \midrule
  Total error (1)+(2) &
    $\eras{+6.2}{-7.6}~\si{\percent} \pm \SI{+7.3}{\percent}$ &
    $\eras{+4.5}{-5.1}~\si{\percent} \pm \SI{+7.3}{\percent}$ \\
  \bottomrule
 \end{tabularx}
\end{table}
}
Systematic errors were evaluated by varying the topological cuts DCA, PA and MM
by \SI{\pm20}{\percent} such as to obtain more strict or more loose selections
as summarised in \cref{tab:cutsvar}. The errors were evaluated independently
for the different phase space bins. Furthermore, the dependence of the
acceptance correction upon the production model was tested by sampling 1000
iterations of the correction matrixes obtained varying each channel \xsec
within the production model according to a Gaussian function with sigma equal
to the channel error. The angular distributions were varied only for two
extreme cases, one assuming the least anisotropic distribution for all the
channels listed in \cref{tab:model} and one assuming the most anisotropic
distributions.
Each channel was sampled independently. For each iteration a new correction
matrix was evaluated and applied to the experimental data. The systematic error
for the acceptance corrections is deduced in each bin from the RMS of the
distribution of reconstructed differential \xsecs using the different
correction matrices.
A summary of all systematical uncertainties is given in \cref{tab:systematic}.
The systematic errors due to the topological cuts reported in
\cref{tab:systematic} are obtained averaging over all the \pcm and \ptcm bins,
respectively. The total errors quoted in \cref{tab:systematic} were obtained by
adding up the topological and acceptance/efficiency correction errors
quadratially.\\
\begin{figure*}[!t]
 \centering
 \begin{subfigure}[b]{0.495\textwidth}
  \includegraphics[width=1.0\textwidth]{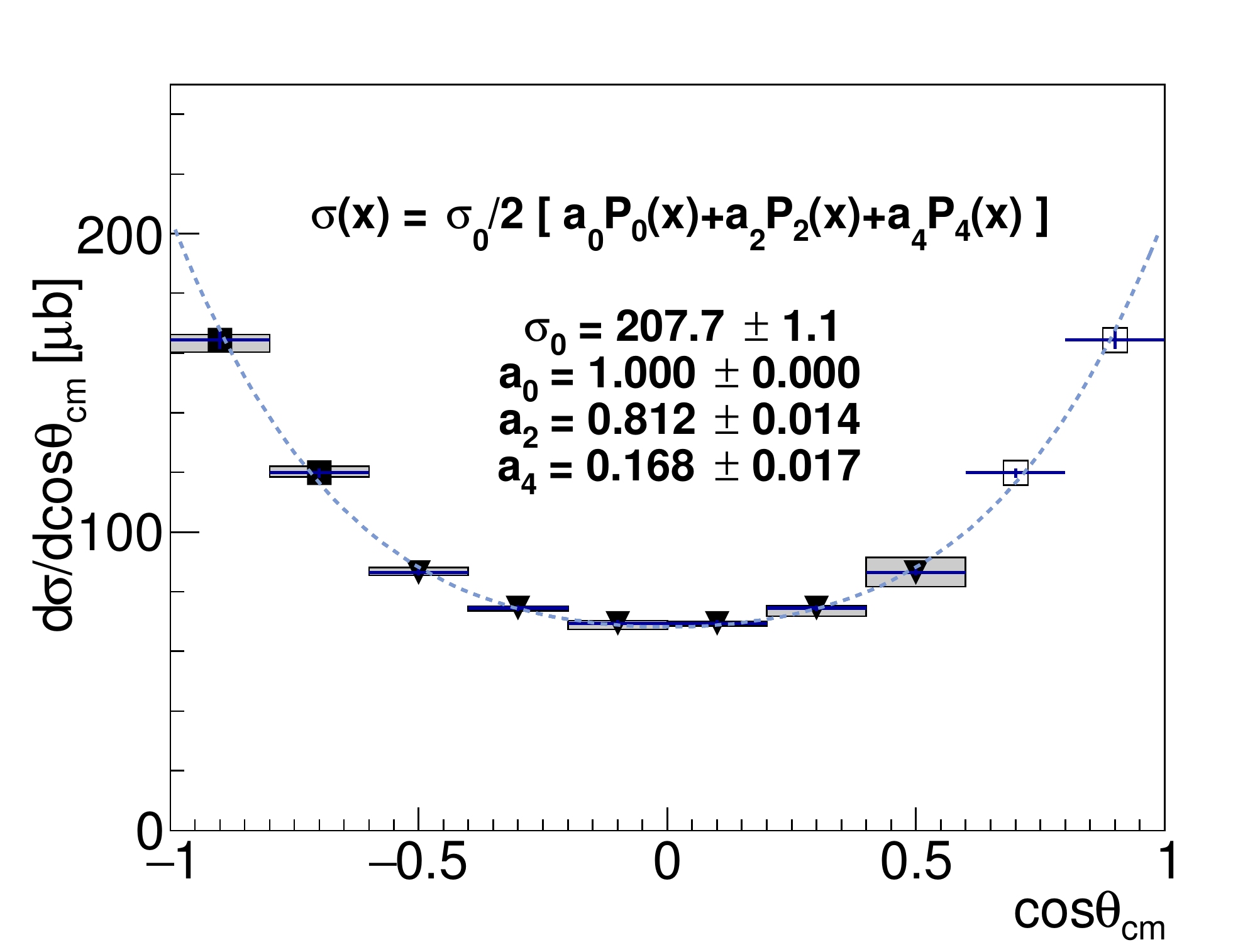}
 \end{subfigure}
%
%
 \begin{subfigure}[b]{0.495\textwidth}
  \includegraphics[width=1.0\textwidth]{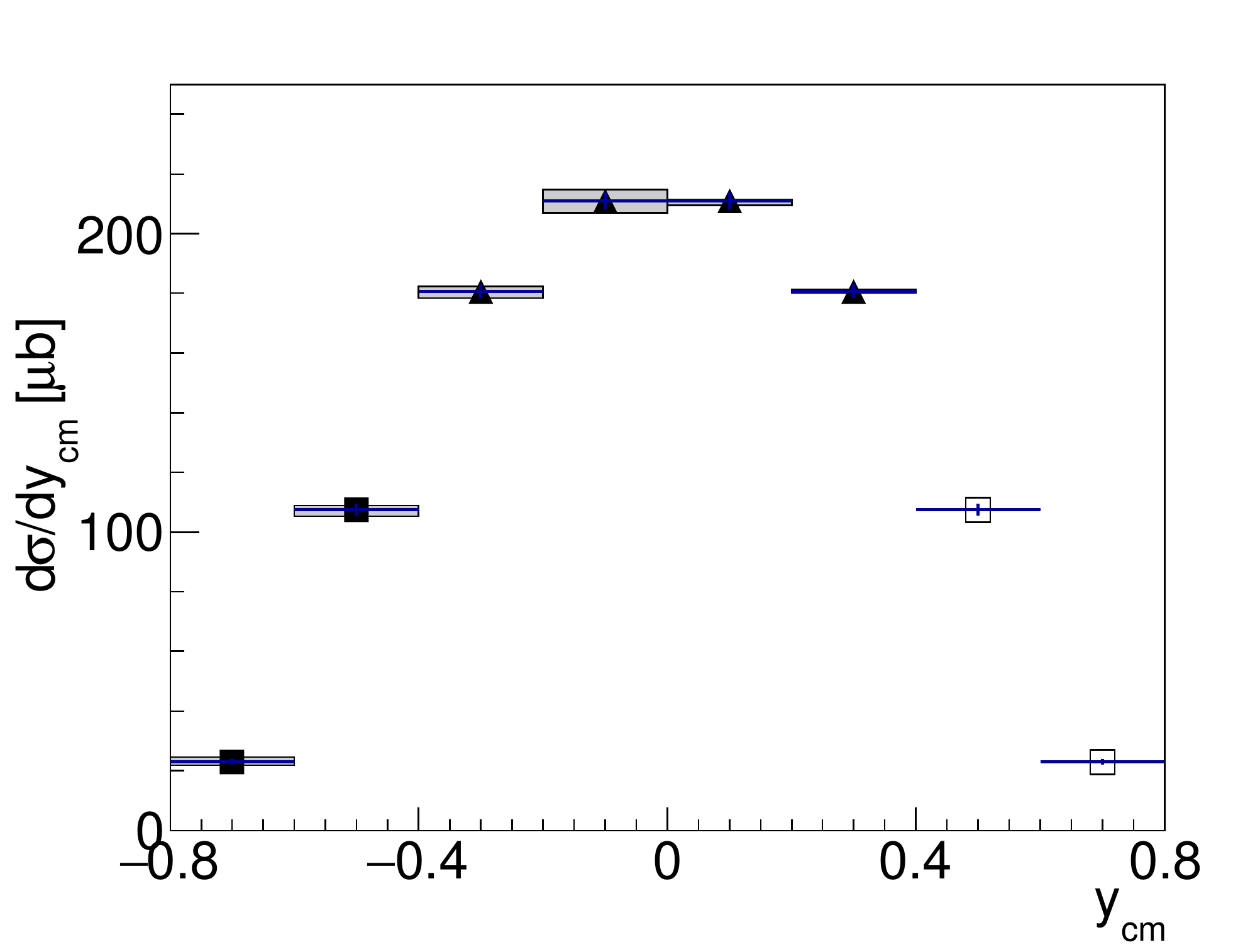}
 \end{subfigure}
 \caption{Differential \xsec distribution $\ud\sigma/\ud$\cthcm (left panel)
    and $\ud\sigma/\ud$\ycm (right panel) extrapolated to the whole phase space
    and integrated over the \pcm and \ptcm variables, respectively. The full
    square symbols represent measured data, the empty squares refer to mirrored
    data and the full triangle correspond to average between mirrored and
    measured data.}
 \label{fig:diffana:exp:cs}
\end{figure*}

\section{Experimental Results and Comparison to the Production Model}
\label{sec4}

\Cref{fig:comp_1,fig:comp_2} show the acceptance and efficiency corrected, and
normalized data as a function of the momentum and polar angle (or transverse
momentum and rapidity) for the $\Lzero$ hyperon calculated in the cms. The
normalization error is equal to \SI{7.28}{\percent}. 

\Cref{fig:comp_1,fig:comp_2} show the comparison of the acceptance and
efficiency corrected experimental data to the total production model, as a
function of the cms momentum \pcm for different polar angle \cthcm bins, and
transverse momentum \ptcm for different rapidity \ycm bins, respectively.
The red symbols depict the corrected experimental data with the systematic
errors shown by the grey boxes.
The normalisation error of \SI{7.28}{\percent} is not shown in
\cref{fig:comp_1,fig:comp_2} but is considered in the evaluation of the error
on the total production \xsec.
The thick black curve shows the resulting distributions from the production
model taking the \xsec and anisotropy parameters listed in \cref{tab:model} and
the error band corresponds to the total errors calculated adding up the \xsec
errors of the different channels as independent errors.
The different contributions to the model are depicted with colour styled
curves.
One can see a good agreement between the production model and the experimental
data for all \cthcm bins except for the most backward direction. One can also
notice that the low momentum range and very forward rapidities are not covered
by the \hades acceptance.\\
The region of $\cos(\theta^\cms) < \num{-0.8}$ shows an enhancement of the
experimental data with respect to the model, especially in the high momentum
region. The sources that mostly contribute in this region of the \phasespace are
the three-body channels $\Lzero\prot\Kp$ and $\Szero\prot\Kp$ (id=1 and 2 in
\cref{tab:model}). Since the $\Szero$ channel was never measured for this beam
energy, we have scaled the \xsec of this channel taking channel 1 as reference
and the prescription from $\hracs$ for the $\Lzero$ to $\Szero$ ratio and
assumed the same anisotropy as for the measurements at lower beam energies
\cite{AbdelBary:2010pc}. Large systematic errors due to both assumptions cannot
be excluded.\\
Possible candidates for the missing yield are channels with double resonance
production (id=\numrange{13}{17}), which were not measured exclusively but their
yield was extracted by fitting simulated data to experimental distributions
\cite{Tsushima:1996xc,PhysRevC.59.369}. In this case, no assumptions could be
made on the corresponding angular distributions that are hence supposed to be
flat. The same problem occurs for the channels 18--23 where the \xsecs could be
estimated but no information about the angular distributions are available.\\
The contribution of the missing higher order final states in the production
model was estimated by fitting the data measured at higher energies 
\cite{lb_v2_rea_1_32} with the phase space parametrisation discussed in
\cite{Agakishiev:2014moo} and extrapolating to the beam energy of this
measurement. The total contribution of these channels was found to be only
\SI{1}{\micro\barn}.
\\
In a second step, the \xsec of the different channels in the production model
were varied to optimize the description of the experimental data. The cocktail
was fit to the experimental data with the \xsecs of all the channels as free
parameters. The boundary for each \xsec was set to the quadratic sum of all the
errors (statistical and systematic) of each channel. The resulting model
obtained after fitting is shown by the red dashed curve in
\cref{fig:comp_1,fig:comp_2}. The corresponding \xsec for each channel is shown 
in the last column of \cref{tab:model}. The error band associated to the result
of the fit represents the total error of the model obtained by summing the \xsec
errors of the different channels.
Since the initial cocktail underestimated the corrected experimental spectra in
the low momentum region, a general increase of the \xsec is expected after the
fit. To compensate the yield increase in the higher \pcm (\ptcm) ranges a few
channels experienced a decrease of the \xsec. All fitted \xsecs reached their
boundary values (denoted as $\top$ and $\bot$ for upper and lower limits,
respectively).

\section{Cross section extraction}
\label{sec5}
\begin{figure}[!t]
 \centering
 \includegraphics[width=1.1\linewidth]{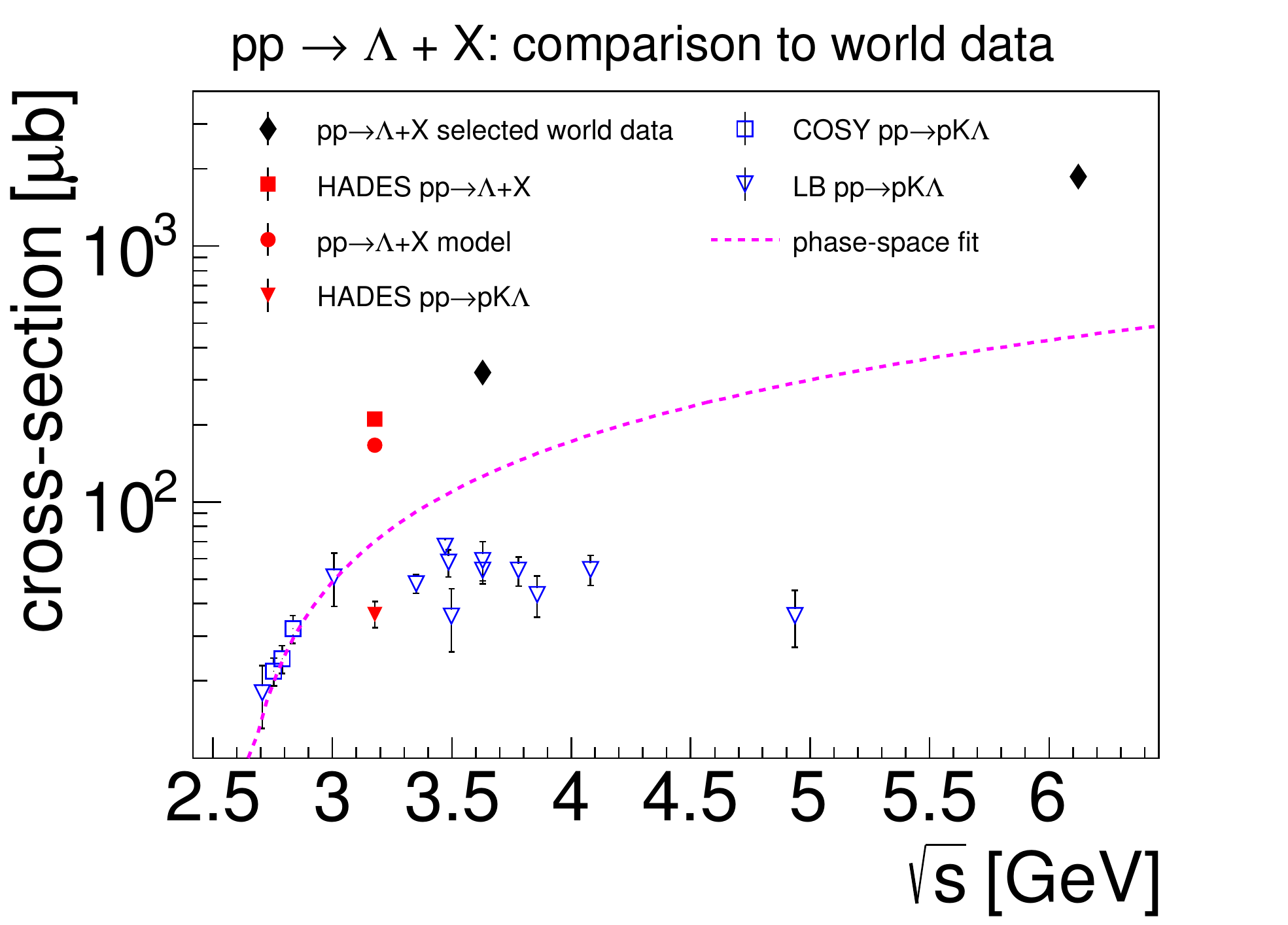}
 \caption{Compilation of the total $\pp\to\Lzero+X$ production \xsec
    measurements. The empty symbols refer to the measurements of the exclusive
    channel $\pp\to\pKL$, the full symbols show the measurements of the
    inclusive $\Lzero$ production and the dashed curve depicts a phase space
    fit for the $\pKL$ exclusive channel in \cite{AbdelBary:2010pc}.}
 \label{exp:wd}
\end{figure}

{
\renewcommand{\arraystretch}{1.4}

\begin{table}[!b]
 \caption[Summary of the \xsec extraction.]{Summary of the extracted \xsec in
    \si{\micro\barn}. The labels EXP and SIM refer to the contribution to the
    total \xsec from the measured values (EXP) and extrapolation of the not
    covered phase space region with simulation (SIM), whereas tot refers to the
    total \xsec. The statistical and systematic errors refer to the experimental
    data and also the total errors on the production model are listed.}
 \label{tab:cs}

 \begin{tabularx}{1.0\linewidth}{lCCCCcC}
  \toprule
  phase space & $\sigma_\mathrm{EXP}$ & $\sigma_\mathrm{SIM}$ &
    $\sigma_\mathrm{tot}$ & $\delta_\mathrm{stat}$ & $\delta_\mathrm{syst}$ &
    $\delta_\mathrm{model}$ \\
  \midrule
  \pcth    & \num{115.0}    & \num{27.8} & \num{205.8} & \num{+-1.5} &
    $\eras{+7.1}{-8.7} \pm \num{8.4}$ & $\eras{+0.3}{-0.4}$ \\
  \pty    & \num{113.8}    & \num{27.0} & \num{208.8} & \num{+-1.1} &
    $\eras{+5.0}{-5.8} \pm \num{8.3}$ & $\eras{+0.5}{-0.6}$ \\
  \midrule
  Total inclusive        & \num{201.2} & --- & \num{201.2} & \num{+-4.1} &
    $\eras{+7.6}{-8.0} \pm \num{14.7}$ & --- \\
  Model                    & --- & \num{165.7} & --- & --- & --- &
    $\eras{+7.6}{-8.2}$ \\
  \bottomrule
 \end{tabularx}
\end{table}
}
The \hades acceptance does not cover the whole \phasespace region, hence
extrapolations must be applied to the experimental data to extract the $4\pi$
yield and the total \xsec. In particular, the spectrometer does not cover the
forward direction (low polar angles, $\mcthcm>\num{0.6}$) and low momenta
($\mpcm / \mptcm < \SI{300}{\mevc}$). 
For the region of $\mpcm / \mptcm< \SI{300}{\mevc}$, the production model is
used to evaluate the $\Lzero$ yield. The \cthcm and \ycm distributions are
obtained integrating the \xsec shown in \cref{fig:comp_1,fig:comp_2} over all
cms momenta and transverse momenta values, respectively.\\
Due to the symmetry of forward and backward directions for the cms variables in
\pp collisions, the \cthcm and \ycm distributions were mirrored at
$\mcthcm = \num{0}$ and $\mycm = 0$, respectively.
If after mirroring two different data points are associated to the same \cthcm
or \ycm bin a new bin content is calculated by the average weighted with the
relative error of each data point.
The error associated to the average was estimated considering the two errors as
independent.\\
The two panels of \cref{fig:diffana:exp:cs} show the differential \xsecs as a
function of the polar angle and rapidity in the cms. The full square symbols
represent the measured data, the empty squares refer to the mirrored data and
the full triangles correspond to the averaged bins between mirrored and
experimental data. For each \phasespace representation a \xsec value was
extracted separately, resulting in 
\eq
\renewcommand{\arraystretch}{2.4}
\begin{array}{l}
 \sigma(\pp\to\Lzero+X) = \num{205.8\pm1.5}\eras{+7.1}{-8.7} \pm \num{8.4}
    \eras{+0.3}{-0.4}~\si{\micro\barn}, \\
 \sigma(\pp\to\Lzero+X) = \num{208.8\pm1.1}\eras{+5.0}{-5.8} \pm \num{8.3}
    \eras{+0.5}{-0.6}~\si{\micro\barn},
\end{array}
\label{eq:cs:diff}
\eeq
for the \cthcm and \ycm distributions, respectively. The \xsecs values were
obtained by considering both the measured and mirrored bins.
\Cref{tab:cs} shows the different contributions to the total \xsecs extracted
from the two phase space representations. The EXP and SIM labels refer to the
contribution to the total \xsec from the measured values (EXP) and the
extrapolation in the phase space region not covered by the HADES geometrical
acceptance via the production model (SIM) before the fit procedure described
in \cref{sec4}. $\sigma_\mathrm{tot}$ represents the total \xsec extracted with
the two differential analyses and the difference between SIM+EXP values and
$\sigma$ stems from the mirrored bins.
One can see that the fraction of the extrapolated yield adds up to about
\SI{14}{\percent} of the total yield and that the mirror-bins contribute with
about \SI{30}{\percent}.
The shown errors refer to the statistical errors, systematic errors due to the
cut variation, normalization error and the model error is equal to the sum of
all the errors on the \xsecs of the different channels.
The final total \xsec was obtained by averaging of both distributions and
is equal to
\eq
 \sigma(\pp\to\Lzero+X) = \num{207.3\pm1.3}\eras{+6.0}{-7.3} \pm \num{8.4} 
    \eras{+0.4}{-0.5}~\si{\micro\barn}.
\label{eq:cs:final}
\eeq
\Cref{tab:cs} contains also the value of the total inclusive $\Lzero$ \xsec
obtained repeating all analysis steps on the whole non-differential data sample
in the same way as described above for the differential analyses. This
cross-check delivers a \xsec value compatible with both the results of the
differential analyses.\\
The \xsec predicted by the production model is also shown in \cref{tab:cs}, it
equals to $\sigma = \num{165.7}\eras{+7.6}{-8.2}~\si{\micro\barn}$, about
\SI{42}{\micro\barn} lower than the experimental value. The \xsec value obtained
after the fit of the different channels to the experimental data amounts to
$\sigma = \num{186.0}\eras{+8.8}{-9.2}~\si{\micro\barn}$.\\
To describe the anisotropy of the total $\Lzero$ distribution in the \cthcm
representation, \cref{eq:csang} was employed to fit the distribution shown in
the left panel of \cref{fig:diffana:exp:cs}. The obtained coefficients $a_2$ and
$a_4$ of the $\Lzero$ anisotropy distribution are listed in \cref{tab:angdist}.
{
\begin{table}[!b]
 \caption{Angular distribution coefficients a$_{0,2,4}$.}
 \label{tab:angdist}

 \begin{tabularx}{1.0\linewidth}{cCCC}
  \toprule
  $\sigma_0$~[\si{\micro\barn}] & $a_0$ & $a_2$ & $a_4$ \\
  \midrule
  \num{207.7 +- 1.1} & \num{1.0} & \num{0.812 +- 0.014} &
    \num{0.168 +- 0.017} \\
  \bottomrule
 \end{tabularx}
\end{table}
}
In \cref{exp:wd}, the \xsec values from our experimental data (red filled
square) and from the model before the fitting (red filled circle) are compared
to the systematics measured previously \cite{Eisner1977361,Alpgard1976349}
(black diamonds) and are found to follow a consistent trend. The exclusive
measurements of the reaction $\pp \rightarrow \pKL$ are also shown in
\cref{exp:wd} by the empty blue symbols \cite{AbdelBary:2010pc,Faldt2014} and
the red full triangle \cite{lb_v2_rea_1_32,Agakishiev2015242}.
This channel is the first contributing to the total $\Lzero$ yield in p+p
reactions and,  while by increasing the beam energy the $\pKL$ \xsec saturates
at a value of the order of \SIrange{30}{40}{\micro\barn}, the \phasespace is 
gradually opened for various  other channels as discussed in this work. The
dashed curve shows the parametrisation by F\"aldt and Wilkin \cite{Faldt2014}
for $\pKL$ \cite{AbdelBary:2010pc}.

\section{Discussion and Summary}
We have measured the inclusive production of $\Lzero$ in p+p collisions at a
kinetic beam energy of \SI{3.5}{\gev} in a fixed target experiment. A total
\xsec of $\num{207.3\pm1.3}\eras{+6.0}{-7.3} \pm \num{8.4}
\eras{+0.4}{-0.5}~\si{\micro\barn}$ is extracted. 
A production model, composed of exclusive channels measured by HADES and
simulations for the non measured production mechanisms, is compared to the
experimental data after efficiency and acceptance corrections for the
reconstruction efficiency and the geometrical acceptance of the spectrometer.
A moderate agreement is found. The differential distributions obtained with the
production model underestimate the measured ones in the region of large
rapidity. The total \xsec from the measurement and from the production model
differs by \SI{20}{\percent}.
If a fit of the simulated channels to the experimental data is performed, by
varying only the production \xsecs within the measured or estimated errors, the
difference between the model and the experimental data decreases to
\SI{15}{\percent}.\\
Beside the lower \xsec, there is also a qualitative difference in the angular
distribution between the experimental data and the production model. This
disagreement could be due to the fact that the \xsec and angular distribution
for the rather abundant $\pKS$ channel were only estimated.\\ 
Other possible candidates for the missing yield are those channels where
$\Lzero$ production is accompanied by a $\Delta$ resonance. Processes where
$\Delta$s and $\Lzero$s are produced directly can be currently estimated only
with the help of models.
Missing higher order final states should contribute with only
\SI{1}{\micro\barn} according to estimations based on the available phase
space.
It is clear that the two crucial aspects are then the direct measurement of the
$\Szero$ channels and the exclusive measurement of the $\Delta$ channels.
The current comparison of the production model and inclusive data allowed to
extract improved \xsec values for the different production channels.\\
The upcoming upgrade of \hades with an Electromagnetic Calorimeter and a Forward
Detector will provide the possibility to measure $\Szero\to\Lzero\gamma$ decays
and reconstruct particles emitted in the very forward direction
($\mcthcm > 0.75$). New measurements of \pp reactions at the same and similar
energies will give the chance to investigate the missing channels and improve
the quality of the production model. Additionally, future $\prot+\prot$
measurements will include the employment of a start detector to improve on the
identification of charged kaons and hence will allow for exclusive measurements
of production channels with $\Delta$ resonances.

\begin{acknowledgments}
The \hades collaboration gratefully acknowledges the support by the grants LIP
Coimbra, Coimbra (Portugal) PTDC/FIS/113339/2009, UJ Kraków (Poland) NCN
2013/10/M/ST2/00042, TU M\"unchen, Garching (Germany) MLL M\"unchen: DFG EClust
153, VH-NG-330 BMBF 06MT9156 TP5 GSI TMKrue 1012 NPI AS CR, Rez, Rez (Czech
Republic) GACR 13-06759S, NPI AS CR, Rez, USC - S. de Compostela, Santiago de
Compostela (Spain) CPAN:CSD2007-00042, Goethe University, Frankfurt (Germany):
HA216/EMMI HIC for FAIR (LOEWE) BMBF:06FY9100I GSI F\&E, IN2P3/CNRS (France).
\end{acknowledgments}


\begin{thebibliography}{36}
\expandafter\ifx\csname natexlab\endcsname\relax\def\natexlab#1{#1}\fi
\expandafter\ifx\csname bibnamefont\endcsname\relax
  \def\bibnamefont#1{#1}\fi
\expandafter\ifx\csname bibfnamefont\endcsname\relax
  \def\bibfnamefont#1{#1}\fi
\expandafter\ifx\csname citenamefont\endcsname\relax
  \def\citenamefont#1{#1}\fi
\expandafter\ifx\csname url\endcsname\relax
  \def\url#1{\texttt{#1}}\fi
\expandafter\ifx\csname urlprefix\endcsname\relax\def\urlprefix{URL }\fi
\providecommand{\bibinfo}[2]{#2}
\providecommand{\eprint}[2][]{\url{#2}}

\bibitem[{\citenamefont{Hartnack et~al.}(2012)\citenamefont{Hartnack, Oeschler,
  Leifels, Bratkovskaya, and Aichelin}}]{Hartnack:2011cn}
\bibinfo{author}{\bibfnamefont{C.}~\bibnamefont{Hartnack}},
  \bibinfo{author}{\bibfnamefont{H.}~\bibnamefont{Oeschler}},
  \bibinfo{author}{\bibfnamefont{Y.}~\bibnamefont{Leifels}},
  \bibinfo{author}{\bibfnamefont{E.~L.} \bibnamefont{Bratkovskaya}},
  \bibnamefont{and} \bibinfo{author}{\bibfnamefont{J.}~\bibnamefont{Aichelin}},
  \bibinfo{journal}{Phys. Rept.} \textbf{\bibinfo{volume}{510}},
  \bibinfo{pages}{119} (\bibinfo{year}{2012}).

\bibitem[{\citenamefont{Lonardoni et~al.}(2015)\citenamefont{Lonardoni, Lovato,
  Gandolfi, and Pederiva}}]{Lonardoni:2014bwa}
\bibinfo{author}{\bibfnamefont{D.}~\bibnamefont{Lonardoni}},
  \bibinfo{author}{\bibfnamefont{A.}~\bibnamefont{Lovato}},
  \bibinfo{author}{\bibfnamefont{S.}~\bibnamefont{Gandolfi}}, \bibnamefont{and}
  \bibinfo{author}{\bibfnamefont{F.}~\bibnamefont{Pederiva}},
  \bibinfo{journal}{Phys. Rev. Lett.} \textbf{\bibinfo{volume}{114}},
  \bibinfo{pages}{092301} (\bibinfo{year}{2015}), \eprint{1407.4448}.

\bibitem[{\citenamefont{Schaffner-Bielich and
  Gal}(2000)}]{SchaffnerBielich:2000wj}
\bibinfo{author}{\bibfnamefont{J.}~\bibnamefont{Schaffner-Bielich}}
  \bibnamefont{and} \bibinfo{author}{\bibfnamefont{A.}~\bibnamefont{Gal}},
  \bibinfo{journal}{Phys. Rev.} \textbf{\bibinfo{volume}{C62}},
  \bibinfo{pages}{034311} (\bibinfo{year}{2000}).

\bibitem[{\citenamefont{Schulze et~al.}(2006)\citenamefont{Schulze, Polls,
  Ramos, and Vidana}}]{Schulze:2006vw}
\bibinfo{author}{\bibfnamefont{H.~J.} \bibnamefont{Schulze}},
  \bibinfo{author}{\bibfnamefont{A.}~\bibnamefont{Polls}},
  \bibinfo{author}{\bibfnamefont{A.}~\bibnamefont{Ramos}}, \bibnamefont{and}
  \bibinfo{author}{\bibfnamefont{I.}~\bibnamefont{Vidana}},
  \bibinfo{journal}{Phys. Rev.} \textbf{\bibinfo{volume}{C73}},
  \bibinfo{pages}{058801} (\bibinfo{year}{2006}).

\bibitem[{\citenamefont{Weissenborn et~al.}(2012)\citenamefont{Weissenborn,
  Chatterjee, and Schaffner-Bielich}}]{Weissenborn:2011kb}
\bibinfo{author}{\bibfnamefont{S.}~\bibnamefont{Weissenborn}},
  \bibinfo{author}{\bibfnamefont{D.}~\bibnamefont{Chatterjee}},
  \bibnamefont{and}
  \bibinfo{author}{\bibfnamefont{J.}~\bibnamefont{Schaffner-Bielich}},
  \bibinfo{journal}{Nucl. Phys.} \textbf{\bibinfo{volume}{A881}},
  \bibinfo{pages}{62} (\bibinfo{year}{2012}), \eprint{1111.6049}.

\bibitem[{\citenamefont{Adamczewski-Musch
  et~al.}(2016)}]{Adamczewski-Musch:2016jlh}
\bibinfo{author}{\bibfnamefont{J.}~\bibnamefont{Adamczewski-Musch}}
  \bibnamefont{et~al.} (\bibinfo{year}{2016}), \eprint{1602.08880}.

\bibitem[{\citenamefont{Li}(2000)}]{Li:1999bea}
\bibinfo{author}{\bibfnamefont{B.-A.} \bibnamefont{Li}},
  \bibinfo{journal}{Phys. Rev.} \textbf{\bibinfo{volume}{C61}},
  \bibinfo{pages}{021903} (\bibinfo{year}{2000}).

\bibitem[{\citenamefont{Fabbietti}(2015)}]{Fabbietti:2015tpa}
\bibinfo{author}{\bibfnamefont{L.}~\bibnamefont{Fabbietti}}
  (\bibinfo{collaboration}{HADES}), in \emph{\bibinfo{booktitle}{{11th
  Conference on Quark Confinement and the Hadron Spectrum (Confinement XI) St.
  Petersburg, Russia, September 8-12, 2014}}} (\bibinfo{year}{2015}).

\bibitem[{\citenamefont{Agakishiev
  et~al.}(2014{\natexlab{a}})}]{Agakishiev:2014kdy}
\bibinfo{author}{\bibfnamefont{G.}~\bibnamefont{Agakishiev}}
  \bibnamefont{et~al.} (\bibinfo{collaboration}{HADES}), \bibinfo{journal}{Eur.
  Phys. J.} \textbf{\bibinfo{volume}{A50}}, \bibinfo{pages}{81}
  (\bibinfo{year}{2014}{\natexlab{a}}), \eprint{1404.3014}.

\bibitem[{\citenamefont{Bass et~al.}(1998)\citenamefont{Bass, Belkacem,
  Bleicher, Brandstetter, Bravina et~al.}}]{Bass:1998ca}
\bibinfo{author}{\bibfnamefont{S.}~\bibnamefont{Bass}},
  \bibinfo{author}{\bibfnamefont{M.}~\bibnamefont{Belkacem}},
  \bibinfo{author}{\bibfnamefont{M.}~\bibnamefont{Bleicher}},
  \bibinfo{author}{\bibfnamefont{M.}~\bibnamefont{Brandstetter}},
  \bibinfo{author}{\bibfnamefont{L.}~\bibnamefont{Bravina}},
  \bibnamefont{et~al.}, \bibinfo{journal}{Prog.Part.Nucl.Phys.}
  \textbf{\bibinfo{volume}{41}}, \bibinfo{pages}{255} (\bibinfo{year}{1998}).

\bibitem[{\citenamefont{Buss et~al.}(2012)\citenamefont{Buss, Gaitanos,
  Gallmeister, van Hees, Kaskulov, Lalakulich, Larionov, Leitner, Weil, and
  Mosel}}]{Buss:2011mx}
\bibinfo{author}{\bibfnamefont{O.}~\bibnamefont{Buss}},
  \bibinfo{author}{\bibfnamefont{T.}~\bibnamefont{Gaitanos}},
  \bibinfo{author}{\bibfnamefont{K.}~\bibnamefont{Gallmeister}},
  \bibinfo{author}{\bibfnamefont{H.}~\bibnamefont{van Hees}},
  \bibinfo{author}{\bibfnamefont{M.}~\bibnamefont{Kaskulov}},
  \bibinfo{author}{\bibfnamefont{O.}~\bibnamefont{Lalakulich}},
  \bibinfo{author}{\bibfnamefont{A.~B.} \bibnamefont{Larionov}},
  \bibinfo{author}{\bibfnamefont{T.}~\bibnamefont{Leitner}},
  \bibinfo{author}{\bibfnamefont{J.}~\bibnamefont{Weil}}, \bibnamefont{and}
  \bibinfo{author}{\bibfnamefont{U.}~\bibnamefont{Mosel}},
  \bibinfo{journal}{Phys. Rept.} \textbf{\bibinfo{volume}{512}},
  \bibinfo{pages}{1} (\bibinfo{year}{2012}), \eprint{1106.1344}.

\bibitem[{\citenamefont{Bastid et~al.}(2007)}]{Bastid:2007jz}
\bibinfo{author}{\bibfnamefont{N.}~\bibnamefont{Bastid}} \bibnamefont{et~al.}
  (\bibinfo{collaboration}{FOPI}), \bibinfo{journal}{Phys. Rev.}
  \textbf{\bibinfo{volume}{C76}}, \bibinfo{pages}{024906}
  (\bibinfo{year}{2007}).

\bibitem[{\citenamefont{Agakishiev
  et~al.}(2014{\natexlab{b}})}]{PhysRevC.90.015202}
\bibinfo{author}{\bibfnamefont{G.}~\bibnamefont{Agakishiev}}
  \bibnamefont{et~al.} (\bibinfo{collaboration}{HADES}),
  \bibinfo{journal}{Phys. Rev.} \textbf{\bibinfo{volume}{C90}},
  \bibinfo{pages}{015202} (\bibinfo{year}{2014}{\natexlab{b}}),
  \eprint{1403.6662}.

\bibitem[{\citenamefont{Agakishiev
  et~al.}(2014{\natexlab{c}})}]{Agakishiev:2014moo}
\bibinfo{author}{\bibfnamefont{G.}~\bibnamefont{Agakishiev}}
  \bibnamefont{et~al.} (\bibinfo{collaboration}{HADES}),
  \bibinfo{journal}{Phys. Rev.} \textbf{\bibinfo{volume}{C90}},
  \bibinfo{pages}{054906} (\bibinfo{year}{2014}{\natexlab{c}}),
  \eprint{1404.7011}.

\bibitem[{\citenamefont{Agakishiev
  et~al.}(2015{\natexlab{a}})}]{Agakishiev:2015ysr}
\bibinfo{author}{\bibfnamefont{G.}~\bibnamefont{Agakishiev}}
  \bibnamefont{et~al.} (\bibinfo{collaboration}{HADES}),
  \bibinfo{journal}{Phys. Rev.} \textbf{\bibinfo{volume}{C92}},
  \bibinfo{pages}{024903} (\bibinfo{year}{2015}{\natexlab{a}}),
  \eprint{1505.06184}.

\bibitem[{\citenamefont{Agakishiev
  et~al.}(2015{\natexlab{b}})}]{Agakishiev2015242}
\bibinfo{author}{\bibfnamefont{G.}~\bibnamefont{Agakishiev}}
  \bibnamefont{et~al.} (\bibinfo{collaboration}{HADES}),
  \bibinfo{journal}{Phys. Lett. B} \textbf{\bibinfo{volume}{742}},
  \bibinfo{pages}{242 } (\bibinfo{year}{2015}{\natexlab{b}}).

\bibitem[{\citenamefont{Agakishiev et~al.}(2012)}]{PhysRevC.85.035203}
\bibinfo{author}{\bibfnamefont{G.}~\bibnamefont{Agakishiev}}
  \bibnamefont{et~al.} (\bibinfo{collaboration}{HADES}),
  \bibinfo{journal}{Phys. Rev.} \textbf{\bibinfo{volume}{C85}},
  \bibinfo{pages}{035203} (\bibinfo{year}{2012}), \eprint{1109.6806}.

\bibitem[{\citenamefont{Agakishiev et~al.}(2013)}]{PhysRevC.87.025201}
\bibinfo{author}{\bibfnamefont{G.}~\bibnamefont{Agakishiev}}
  \bibnamefont{et~al.} (\bibinfo{collaboration}{HADES}),
  \bibinfo{journal}{Phys. Rev.} \textbf{\bibinfo{volume}{C87}},
  \bibinfo{pages}{025201} (\bibinfo{year}{2013}), \eprint{1208.0205}.

\bibitem[{\citenamefont{Siebenson and Fabbietti}(2013)}]{Siebenson:2013rpa}
\bibinfo{author}{\bibfnamefont{J.}~\bibnamefont{Siebenson}} \bibnamefont{and}
  \bibinfo{author}{\bibfnamefont{L.}~\bibnamefont{Fabbietti}},
  \bibinfo{journal}{Phys. Rev.} \textbf{\bibinfo{volume}{C88}},
  \bibinfo{pages}{055201} (\bibinfo{year}{2013}), \eprint{1306.5183}.

\bibitem[{\citenamefont{Agakishiev
  et~al.}(2015{\natexlab{c}})}]{Agakishiev:2014dha}
\bibinfo{author}{\bibfnamefont{G.}~\bibnamefont{Agakishiev}}
  \bibnamefont{et~al.} (\bibinfo{collaboration}{HADES}),
  \bibinfo{journal}{Phys. Lett.} \textbf{\bibinfo{volume}{B742}},
  \bibinfo{pages}{242} (\bibinfo{year}{2015}{\natexlab{c}}).

\bibitem[{\citenamefont{Abdel-Bary et~al.}(2010)}]{AbdelBary:2010pc}
\bibinfo{author}{\bibfnamefont{M.}~\bibnamefont{Abdel-Bary}}
  \bibnamefont{et~al.} (\bibinfo{collaboration}{COSY-TOF}),
  \bibinfo{journal}{Eur. Phys. J.} \textbf{\bibinfo{volume}{A46}},
  \bibinfo{pages}{27} (\bibinfo{year}{2010}), \bibinfo{note}{[Erratum: Eur.
  Phys. J.A46,435(2010)]}.

\bibitem[{\citenamefont{Fabbietti and Epple}(2010)}]{Fabbietti2010333}
\bibinfo{author}{\bibfnamefont{L.}~\bibnamefont{Fabbietti}} \bibnamefont{and}
  \bibinfo{author}{\bibfnamefont{E.}~\bibnamefont{Epple}},
  \bibinfo{journal}{Nucl. Phys.} \textbf{\bibinfo{volume}{A835}},
  \bibinfo{pages}{333} (\bibinfo{year}{2010}), \bibinfo{note}{proceedings of
  the 10th International Conference on Hypernuclear and Strange Particle
  Physics}.

\bibitem[{\citenamefont{Agakishiev
  et~al.}(2014{\natexlab{d}})}]{PhysRevC.90.054906}
\bibinfo{author}{\bibfnamefont{G.}~\bibnamefont{Agakishiev}}
  \bibnamefont{et~al.} (\bibinfo{collaboration}{HADES}),
  \bibinfo{journal}{Phys. Rev.} \textbf{\bibinfo{volume}{C90}},
  \bibinfo{pages}{054906} (\bibinfo{year}{2014}{\natexlab{d}}),
  \eprint{1404.7011}.

\bibitem[{\citenamefont{Agakishiev et~al.}(2009)}]{Agakishiev:2009am}
\bibinfo{author}{\bibfnamefont{G.}~\bibnamefont{Agakishiev}}
  \bibnamefont{et~al.} (\bibinfo{collaboration}{HADES}), \bibinfo{journal}{Eur.
  Phys. J.} \textbf{\bibinfo{volume}{A41}}, \bibinfo{pages}{243}
  (\bibinfo{year}{2009}), \eprint{0902.3478}.

\bibitem[{\citenamefont{Olive et~al.}(2014)}]{Agashe:2014kda}
\bibinfo{author}{\bibfnamefont{K.~A.} \bibnamefont{Olive}} \bibnamefont{et~al.}
  (\bibinfo{collaboration}{Particle Data Group}), \bibinfo{journal}{Chin.
  Phys.} \textbf{\bibinfo{volume}{C38}}, \bibinfo{pages}{090001}
  (\bibinfo{year}{2014}).

\bibitem[{\citenamefont{Lalik}(2015)}]{PoS_Bormio2015_009}
\bibinfo{author}{\bibfnamefont{R.}~\bibnamefont{Lalik}}, in
  \emph{\bibinfo{booktitle}{{53rd International Winter Meeting on Nuclear
  Physics}}} (\bibinfo{publisher}{PoS}, \bibinfo{year}{2015}), vol.
  \bibinfo{volume}{PoS(Bormio2015)009}.

\bibitem[{\citenamefont{Tsushima et~al.}(1997)\citenamefont{Tsushima,
  Sibirtsev, and Thomas}}]{Tsushima:1996xc}
\bibinfo{author}{\bibfnamefont{K.}~\bibnamefont{Tsushima}},
  \bibinfo{author}{\bibfnamefont{A.}~\bibnamefont{Sibirtsev}},
  \bibnamefont{and} \bibinfo{author}{\bibfnamefont{A.~W.}
  \bibnamefont{Thomas}}, \bibinfo{journal}{Phys. Lett.}
  \textbf{\bibinfo{volume}{B390}}, \bibinfo{pages}{29} (\bibinfo{year}{1997}),
  ISSN \bibinfo{issn}{0370-2693}.

\bibitem[{\citenamefont{Tsushima et~al.}(1999)\citenamefont{Tsushima,
  Sibirtsev, Thomas, and Li}}]{PhysRevC.59.369}
\bibinfo{author}{\bibfnamefont{K.}~\bibnamefont{Tsushima}},
  \bibinfo{author}{\bibfnamefont{A.}~\bibnamefont{Sibirtsev}},
  \bibinfo{author}{\bibfnamefont{A.~W.} \bibnamefont{Thomas}},
  \bibnamefont{and} \bibinfo{author}{\bibfnamefont{G.~Q.} \bibnamefont{Li}},
  \bibinfo{journal}{Phys. Rev.} \textbf{\bibinfo{volume}{C59}},
  \bibinfo{pages}{369} (\bibinfo{year}{1999}), \bibinfo{note}{[Erratum: Phys.
  Rev.C61,029903(2000)]}.

\bibitem[{\citenamefont{Fabbietti et~al.}(2013)}]{Fabbietti:2013npl}
\bibinfo{author}{\bibfnamefont{L.}~\bibnamefont{Fabbietti}}
  \bibnamefont{et~al.}, \bibinfo{journal}{Nucl. Phys.}
  \textbf{\bibinfo{volume}{A914}}, \bibinfo{pages}{60} (\bibinfo{year}{2013}).

\bibitem[{\citenamefont{Baldini et~al.}(1988)\citenamefont{Baldini, Flaminio,
  Moorhead, and Morrison}}]{lb_v2_rea_1_32}
\bibinfo{author}{\bibfnamefont{A.}~\bibnamefont{Baldini}},
  \bibinfo{author}{\bibfnamefont{V.}~\bibnamefont{Flaminio}},
  \bibinfo{author}{\bibfnamefont{W.~G.} \bibnamefont{Moorhead}},
  \bibnamefont{and} \bibinfo{author}{\bibfnamefont{D.~R.~O.}
  \bibnamefont{Morrison}} (\bibinfo{year}{1988}), vol. \bibinfo{volume}{12b} of
  \emph{\bibinfo{series}{{Landolt-B\"{o}rnstein - Group I Elementary Particles,
  Nuclei and Atoms}}}, chap. \bibinfo{chapter}{Reaction 1 - 32}.

\bibitem[{\citenamefont{Frohlich et~al.}(2007)}]{Frohlich:2007bi}
\bibinfo{author}{\bibfnamefont{I.}~\bibnamefont{Frohlich}}
  \bibnamefont{et~al.}, \bibinfo{journal}{PoS}
  \textbf{\bibinfo{volume}{ACAT2007}}, \bibinfo{pages}{076}
  (\bibinfo{year}{2007}), \eprint{0708.2382}.

\bibitem[{\citenamefont{Brun et~al.}(1978)\citenamefont{Brun, Hagelberg,
  Hansroul, and Lassalle}}]{Brun:118715}
\bibinfo{author}{\bibfnamefont{R.}~\bibnamefont{Brun}},
  \bibinfo{author}{\bibfnamefont{R.}~\bibnamefont{Hagelberg}},
  \bibinfo{author}{\bibfnamefont{M.}~\bibnamefont{Hansroul}}, \bibnamefont{and}
  \bibinfo{author}{\bibfnamefont{J.~C.} \bibnamefont{Lassalle}},
  \emph{\bibinfo{title}{{Simulation program for particle physics experiments,
  GEANT: user guide and reference manual}}} (\bibinfo{publisher}{CERN},
  \bibinfo{address}{Geneva}, \bibinfo{year}{1978}).

\bibitem[{\citenamefont{Rustamov and
  Collaboration}(2010)}]{aipcp:10.1063:1.3483432}
\bibinfo{author}{\bibfnamefont{A.}~\bibnamefont{Rustamov}} \bibnamefont{and}
  \bibinfo{author}{\bibfnamefont{T.~H.} \bibnamefont{Collaboration}},
  \bibinfo{journal}{AIP Conference Proceedings}
  \textbf{\bibinfo{volume}{1257}}, \bibinfo{pages}{736–740}
  (\bibinfo{year}{2010}).

\bibitem[{\citenamefont{Eisner et~al.}(1977)\citenamefont{Eisner, Fickinger,
  Glickman, Malko, Owens, Robinson, Dado, Engler, Keyes, Kikuchi
  et~al.}}]{Eisner1977361}
\bibinfo{author}{\bibfnamefont{R.}~\bibnamefont{Eisner}},
  \bibinfo{author}{\bibfnamefont{W.}~\bibnamefont{Fickinger}},
  \bibinfo{author}{\bibfnamefont{S.}~\bibnamefont{Glickman}},
  \bibinfo{author}{\bibfnamefont{J.}~\bibnamefont{Malko}},
  \bibinfo{author}{\bibfnamefont{J.}~\bibnamefont{Owens}},
  \bibinfo{author}{\bibfnamefont{D.}~\bibnamefont{Robinson}},
  \bibinfo{author}{\bibfnamefont{S.}~\bibnamefont{Dado}},
  \bibinfo{author}{\bibfnamefont{A.}~\bibnamefont{Engler}},
  \bibinfo{author}{\bibfnamefont{G.}~\bibnamefont{Keyes}},
  \bibinfo{author}{\bibfnamefont{T.}~\bibnamefont{Kikuchi}},
  \bibnamefont{et~al.}, \bibinfo{journal}{Nucl. Phys.}
  \textbf{\bibinfo{volume}{B123}}, \bibinfo{pages}{361} (\bibinfo{year}{1977}).

\bibitem[{\citenamefont{Alpg{\aa}rd et~al.}(1976)\citenamefont{Alpg{\aa}rd,
  Andersen, Frodesen, Hagman, Hulth, Svedin, Tuominiemi, Villanen, and
  Yamdagni}}]{Alpgard1976349}
\bibinfo{author}{\bibfnamefont{K.}~\bibnamefont{Alpg{\aa}rd}},
  \bibinfo{author}{\bibfnamefont{V.}~\bibnamefont{Andersen}},
  \bibinfo{author}{\bibfnamefont{A.}~\bibnamefont{Frodesen}},
  \bibinfo{author}{\bibfnamefont{V.-M.} \bibnamefont{Hagman}},
  \bibinfo{author}{\bibfnamefont{P.}~\bibnamefont{Hulth}},
  \bibinfo{author}{\bibfnamefont{U.}~\bibnamefont{Svedin}},
  \bibinfo{author}{\bibfnamefont{J.}~\bibnamefont{Tuominiemi}},
  \bibinfo{author}{\bibfnamefont{P.}~\bibnamefont{Villanen}}, \bibnamefont{and}
  \bibinfo{author}{\bibfnamefont{N.}~\bibnamefont{Yamdagni}},
  \bibinfo{journal}{Nucl. Phys.} \textbf{\bibinfo{volume}{B105}},
  \bibinfo{pages}{349} (\bibinfo{year}{1976}).

\bibitem[{\citenamefont{F\"{a}ldt and Wilkin}(2014)}]{Faldt2014}
\bibinfo{author}{\bibfnamefont{G.}~\bibnamefont{F\"{a}ldt}} \bibnamefont{and}
  \bibinfo{author}{\bibfnamefont{C.}~\bibnamefont{Wilkin}},
  \bibinfo{journal}{Zeitschrift f\"{u}r Physik A Hadrons and Nuclei}
  \textbf{\bibinfo{volume}{357}}, \bibinfo{pages}{241} (\bibinfo{year}{2014}).

\end{thebibliography}
\end{document}